\documentclass[12pt,preprint,apj]{emulateapj}
\newcommand{\kmsMpc}{{km s$^{-1}$ Mpc$^{-1}$}}

\usepackage{epstopdf}
\usepackage{multirow}
\usepackage{color}
\shorttitle{A Revised TRGB Calibration}
\shortauthors{Jang \& Lee}

\begin{document}

\title{
%{\bf DRAFT  \today}} \\
The Tip of the Red Giant Branch Distances to Type Ia Supernova Host Galaxies. IV. Color Dependence and Zero-Point Calibration}

\author{In Sung Jang  and  Myung Gyoon Lee}
\affil{Astronomy Program, Department of Physics and Astronomy, Seoul National University, Gwanak-gu, Seoul 151-742, Korea}
\email{isjang@astro.snu.ac.kr, mglee@astro.snu.ac.kr }

%==============================================================================================================

\begin{abstract}
%\noindent
We present a revised TRGB calibration, accurate to 2.7\% of distance.
A modified TRGB magnitude corrected for the color dependence of the TRGB, the $QT$ magnitude, is introduced for better measurement of the TRGB.
We determine the color-magnitude relation of the TRGB from photometry of deep images of HST/ACS fields around eight nearby galaxies. 
The zero-point of the TRGB at the fiducial metallicity ([Fe/H]$=-1.6$ ($(V-I)_{0,TRGB}=1.5$)) is obtained from  photometry of two distance anchors, NGC 4258 (M106) and the LMC, to which precise geometric distances are known: $M_{QT,TRGB}=-4.023\pm0.073$ mag from NGC 4258 and $M_{QT,TRGB}=-4.004\pm0.096$ mag from the LMC.
A weighted mean of the two zero-points is  $M_{QT,TRGB}=-4.016\pm0.058$ mag. 
Quoted uncertainty is $\sim2$ times smaller than those of the previous calibrations.
We compare the empirical TRGB calibration derived in this study with theoretical stellar models, finding that there are significant discrepancies, especially for red  color ($(\rm F606W-F814W)_0\gtrsim2.5$).
We provide the revised TRGB calibration in several magnitude systems for future studies.
\end{abstract}


\section{Introduction}

The Tip of the Red Giant Branch (TRGB) represents the brightest part of the Red Giant Branch (RGB) in the color-magnitude diagrams (CMDs) of old stellar systems such as globular clusters and halos in galaxies. It corresponds to the core-He flash point in the evolutionary stages of low mass stars.
An implication of the TRGB as a distance indicator was suggested first by \citet{baa44}. In his pioneering paper introducing the concept of stellar populations I and II, he investigated the red-sensitive plates of three early type stellar systems, M32, NGC 205, and the central region (bulge) of M31, concluding that the brightest stars (RGB stars) in these three systems have similar magnitudes and colors.
\citet{san71} pointed out that photographs of the Local Group galaxies invariably show surrounding red stars, which have similar absolute magnitudes and colors. Based on the Cepheid distances to three Local Group galaxies (M31, M33 and IC 1613), he derived a mean absolute magnitude of these brightest red stars to be $M_V=-3.0\pm0.2$ mag.

\citet{mou83, mou84} and \citet{mou86} carried out CCD photometry of four Local Group galaxies (M31, M33, NGC 147 and NGC 205) and determined the TRGB distances to these galaxies, which are not much different from previous distance estimates to M31 and M33 based on Cepheid variables.
They adopted a bolometric magnitude of the TRGB ($M_{\rm bol}\sim-3.5$ mag) given by \citet{fro83}.
\citet{fre88b} presented CCD photometry of two fields in IC 1613. Color-magnitude diagrams (CMDs) of these fields show a clear RGB population and the TRGB. Adopting the bolometric magnitude the TRGB ($M_{\rm bol}\sim-3.5$ mag), she derived remarkably similar TRGB distances between the two fields, showing only 0.04 mag difference. Moreover, the TRGB distance from her photometry is in excellent agreement with the Cepheid distance to IC 1613 obtained by \citet{fre88a}.

%%%%%%%%%%%%%%%%%%%%%%%%%%%%%%%%%%%%%%%%%%%
%% TABLE 1
%%%%%%%%%%%%%%%%%%%%%%%%%%%%%%%%%
\begin{deluxetable*}{lcrrrrrrr}
\tabletypesize{\footnotesize}
%\tabletypesize{\scriptsize}
%\tabletypesize{\tiny}
\setlength{\tabcolsep}{0.05in}
%\rotate
\tablecaption{A Summary of $HST$ Observations of the TRGB Calibration Sample}
\tablewidth{0pt}
\tablehead{ \colhead{Target} & \colhead{Field} & \colhead{R.A.} &  \colhead{Decl.} & \colhead{Instrument} & \multicolumn{3}{c}{Exposure time} & \colhead{Prop. ID} \\
& & (J2000.0) & (J2000.0) & & F555W & F606W & F814W
}
\startdata
M105 	& F1 & 10 48 01.53 	& 12 32 28.2 & ACS/WFC & & 9,775s & 9,775s & 10413 \\
	 	& F2 & 10 48 00.20 	& 12 32 06.8 & ACS/WFC & & 9,775s & 9,775s & 10413 \\
	 	
NGC 3384& F1 & 10 48 26.03	& 12 39 50.6 & ACS/WFC & & 14,575s & 14,575s & 10413 \\
	 	& F2 & 10 48 22.38	& 12 39 08.6 & ACS/WFC & & 9,825s & 9,825s & 10413 \\
	 		 	
M81		& F1 & 09 54 34.56	& 69 16 50.9 & ACS/WFC & & 24,232s& 29,953s& 10915	\\
		& F2 & 09 54 13.96	& 69 05 42.6 & ACS/WFC & & 5,354s & 5,501s & 10136 \\
		& F3 & 09 57 01.20	& 68 55 55.6 & ACS/WFC & & 1,580s & 1,595s & 10584 \\
NGC 3377& F1 & 10 47 48.11 	& 13 55 45.2 & ACS/WFC & & 38,500s & 22,260s & 9811 \\
NGC 253 & F1 & 00 48 19.46	&--25 08 47.4& ACS/WFC & & 2,283s & 2,253s & 10915 \\
NGC 4258& F1 & 12 19 18.37	& 47 20 13.8 & ACS/WFC & 5,700s & & 2,600s & 9477 \\	
NGC 300 & F1 & 00 54 47.98	&--37 40 51.6& ACS/WFC & & 1,515s & 1,542s & 10915\\
		& F2 & 00 54 34.70	&--37 39 25.4& ACS/WFC & & 1,515s & 1,542s & 10915\\
		& F3 & 00 54 21.37	&--37 37 56.3& ACS/WFC & & 1,515s & 1,542s & 10915\\
NGC 3077& F1 & 10 03 28.41	& 68 43 52.8 & ACS/WFC & & 1,596s & 1,622s & 10915 \\
\hline
\enddata
\label{observation}
\end{deluxetable*}

A feasibility of the TRGB as a robust distance indicator  was established by \citet{lee93}. 
They investigated empirical RGB loci of the Milky Way globular clusters \citep{dac90} and Yale theoretical stellar isochrones \citep{gre87}, finding that the $I$-band magnitude of the TRGB is almost constant with variation smaller than 0.1 mag ($M_{I,{\rm TRGB}} = -4.0\pm0.1$)
for low metallicity ([Fe/H]$\leq-0.7$) or for
age older than 3 Gyr.
Also, they applied a Sobel kernel for detecting the TRGB from the discontinuity of the $I$-band luminosity function.
They found that a comparison of the TRGB distances to 12 Local Group galaxies with the distances based on Cepheid and RR-Lyrae variables shows a good agreement, showing that the precision of the TRGB is comparable to these primary distance indicators. 
With the TRGB, distances to more than three hundreds of nearby galaxies have been measured \citep{fre10,dal09,rad11,tul13,lee13,lee16}.

In comparison with the Cepheid variables, which are a well known primary distance indicator, the TRGB has several advantages.
First, the TRGB method uses Population II RGB stars. It enables to measure distances to both early type galaxies and late type galaxies, while Cepheid variables can be used only for late type galaxies.
Second, the TRGB method mainly uses RGB stars in the halo regions, where stellar densities are low and internal extinctions are negligible, while Cepheid variables are located in the star-forming regions where internal extinctions are often significant.
Third, the TRGB stars  are non-variable stars, so that a single epoch observation is enough to get distances, while Cepheids require many epochs of observations.
Fourth, the metallicity dependence of the $I$-band magnitude of the TRGB is known to be small for low metallicity\citep{lee93, riz07, mad09, bel08}, while the metallicity effect on the optical luminosity of Cepheids is controversial \citep{ger11}.

On the other hand, the TRGB has a disadvantage that its luminosity is relatively fainter than those of the bright Cepheid variables. 
The absolute magnitude of the brightest Cepheid variable is estimated to be $M_V\sim-7$ mag, about 4.5 mag brighter than that of the TRGB, $M_V\sim-2.5$ mag. 
This difference becomes smaller in the longer wavelength: 
$\Delta \rm mag=3.9$, 3.1, 2.6 and 2.6 mag in $I$, $J$, $H$, and $K_s$ bands, respectively \citep[][assuming the maximum period of Cepheid variables $(P_{max}$ = 100 days)]{mac06, mac15}.

%Taking advantage of the TRGB, we started a project to improve luminosity calibration of Type Ia supernovae (SNe Ia) by measuring accurate TRGB distances to SN Ia host galaxies.  
Taking advantage of the TRGB, we started a project, the TRGB distances to SN host galaxies in the Universe ({\color{red} \bf TIPSNU}), to improve luminosity calibration of SNe.
\citet{lee12} (Paper I) presented a TRGB distance to nearby spiral galaxy M101 hosting SN 2011fe.
\citet{lee13} (Paper II) determined  TRGB distances to M66 and M96 in the Leo I group hosting two SNe Ia, SN 1989B and SN 1998bu. 
Later, \citet{jan15} (Paper III) obtained TRGB distances to two galaxies, NGC 4038/39 hosting SN 2007sr and NGC 5584 hosting SN 2007af.  Combining the TRGB distances and optical light curves of the five SNe Ia, \citet{jan15} derived a value of the Hubble constant, $H_0 = 69.8\pm2.6\pm3.9$ \kmsMpc. This value is similar to those from the cosmic microwave background \citep{ben14,pla15} but still agrees with in errors from those of Cepheid calibrated SNe Ia \citep{rie11,fre12, rie16}.
In order to make better estimation of $H_0$, it is needed to reduce both random and systematic uncertainties.

This paper, the fourth paper of our series, is aiming to refine two issues in the TRGB method: a calibration of the color (metallicity) -- luminosity relation and a calibration of the zero-point of the TRGB magnitude. 
Both calibrations are important to reduce the systematic uncertainty of the TRGB distance estimates and of the value of the Hubble constant.
Previous color dependence calibrations were based on theoretical stellar models \citep{sal97,fus12},
or on a small number of 
stellar systems (Milky Way globular clusters or nearby galaxies) showing a narrow RGB color range \citep{lee93,bel01,bel04, riz07, bel08, mad09}. These calibrations show a significant discrepancy of the color -- luminosity relation ranging from flat to steep (as shown in the Discussion section). 

Zero-point calibrations of the TRGB in previous studies were done with theoretical models \citep{sal97,fus12} or observational approaches \citep{lee93, fer00, bel01,bel04, riz07, bel08, tam08, mad09}.
Various distance indicators (RR Lyrae stars, Cepheids, eclipsing binary stars, and horizontal branch stars) have been used for this calibration. However, derived values show a non-negligible scatter with a large uncertainty,
as shown in the Discussion section.

This paper is composed of as follows. 
In Section 2 we describe how we derive color (metallicity) - luminosity relation of the TRGB. 
Target galaxy selection criteria and data reduction methods will be presented. \S 3 shows the zero-point calibration of the TRGB based on two distance anchors: NGC 4258 and the LMC. A summary of the revised TRGB calibration including various filter systems is shown in \S 4. 
We discuss implications of our results in \S 5 and summarize primary results in the final section.

\section{Color Dependence Calibration}

% Figure 1
\begin{figure} 
\centering
\includegraphics[scale=0.8]{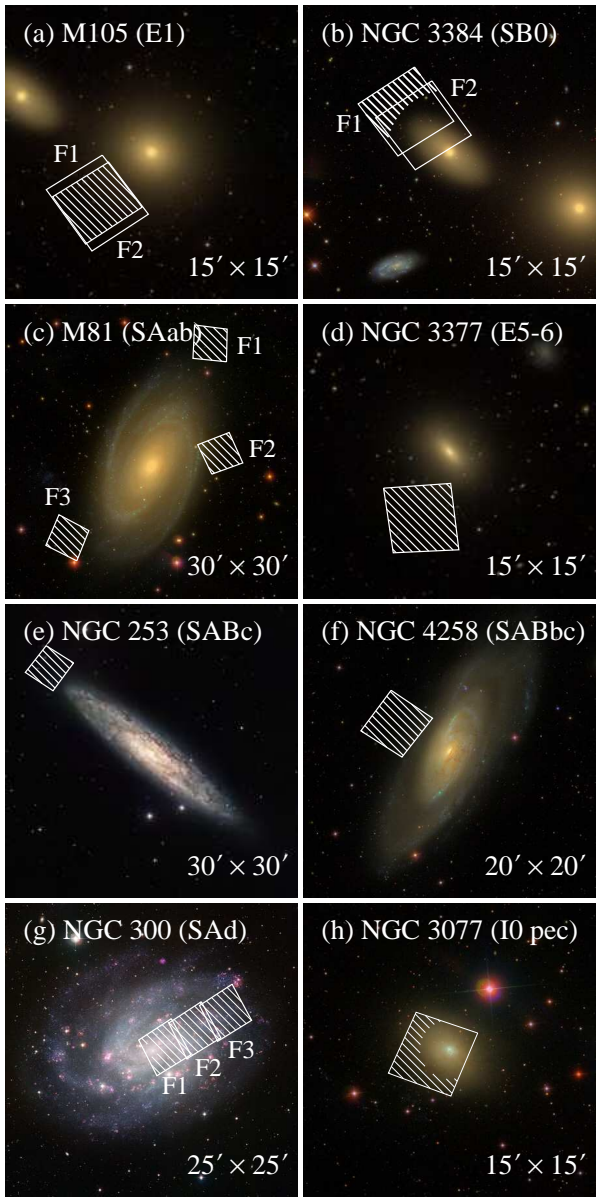} %white.eps}
\caption{
Finding charts for eight galaxies used  for the color dependence calibration of the TRGB: M105 (a), NGC 3384 (b), M81 (c), NGC 3377 (d), NGC 253 (e), NGC 4258 (f), NGC 300 (g), and NGC 3077 (h). HST/ACS fields used in this study are marked on the color maps of the Sloan Digital Sky Survey (SDSS) or public images provided by the European Southern Observatory (ESO). 
The regions we used for the TRGB calibration are indicated by the hatched regions.
}
\label{fig_finder}
\end{figure}

\subsection{Target Selection and Data Reduction}

Optimal target galaxies and target fields for the color (metallicity) dependence  calibration of the TRGB are expected to satisfy the following criteria: 
1) showing a wide color range of the TRGB, covering from the metal poor (blue) RGB stars to the metal rich (red) RGB stars, 
2) allowing a high signal to noise detection of RGB stars, 
3) showing a dominant old RGB population, 
and
4) being observed with the same instrument to reduce any chance for increasing instrument-dependent uncertainties (e.g. photometric transformation uncertainties).
Considering these selection criteria, 
we selected 14 fields around 8 nearby galaxies for which $HST$/Advanced Camera for Surveys ($ACS$) images are available in the archive:  M105 (E1), NGC 3384 (SB0), M81 (SA(s)AB), NGC 3377 (E5-6), NGC 253 (SAB(s)c), NGC 4258 (SAB(s)bc), NGC 300 (SA(s)d), and NGC 3077 (I0 pec).

These galaxies have various morphological types from elliptical galaxies to spiral and irregular galaxies.
Identifications of the $HST$ fields in each galaxy are shown in {\color{red}{\bf Figure \ref{fig_finder}}}.
These $HST$ fields are located from the center to outer regions in the target galaxies.  
Note that the three fields in NGC 300 are located in the disk.
\citet{riz07} used two $HST/ACS$ fields in other positions of the NGC 300 disk for their TRGB color calibration.
For this reason, we also included disk fields of NGC 300 for which deeper HST/ACS images are available. 
{\color{red}{\bf Table \ref{observation} }} lists a summary of $HST/ACS$ observations of the target galaxies. Exposure times are ranging from 1,515s to 38,500s, long enough to detect resolved red giants.
It is noted that NGC 4258 was observed with F555W filter, while other galaxies were observed with F606W filter. 

We obtained ACS images of the fields 
from the $HST$ archive and constructed deep master drizzled images, following the procedures described in \citet{jan15}. In our previous studies \citep{lee12, lee13, jan15}, we carried out point spread function (PSF) photometry using DAOPHOT in IRAF package \citep{ste87}. 
We noted, however, that most of the previous and on-going studies for the calibration of SNe Ia (e.g. the Hubble Key Project \citep{fre01}, the SN HST project \citep{san06}, the SH0ES Project \citep{rie11}, and the Carnegie-Chicago Hubble Program \citep{bea16}) have been done 
with the standalone version of DAOPHOT \citep{ste87}.  
Moreover, the Optical Gravitational Lensing Experiment (OGLE) photometric data of the LMC \citep{ula12}, which we used in the zero-point calibration of the TRGB (see Section 3.2) is also based on the standalone version of DAOPHOT.
Thus, we use the standalone version of DAOPHOT in this study, to reduce software dependent uncertainty as much as possible. Our previous results based on IRAF/DAOPHOT will be updated in the upcoming paper (Jang \& Lee 2016, in preparation).
A single pass sequence (DAOFIND, PHOT, ALLSTAR and ALLFRAME) of 
PSF photometry  was carried out. 
We used PSF images produced from isolated bright stars in each $ACS$ field. We adopted updated photometric zero-points and aperture correction values provided by the STScI webpage\footnote{http://www.stsci.edu/hst/acs/analysis/zeropoints}.

\subsection{CMDs of Resolved Stars and the TRGB detection}

\begin{figure*}
\centering
\includegraphics[scale=0.55]{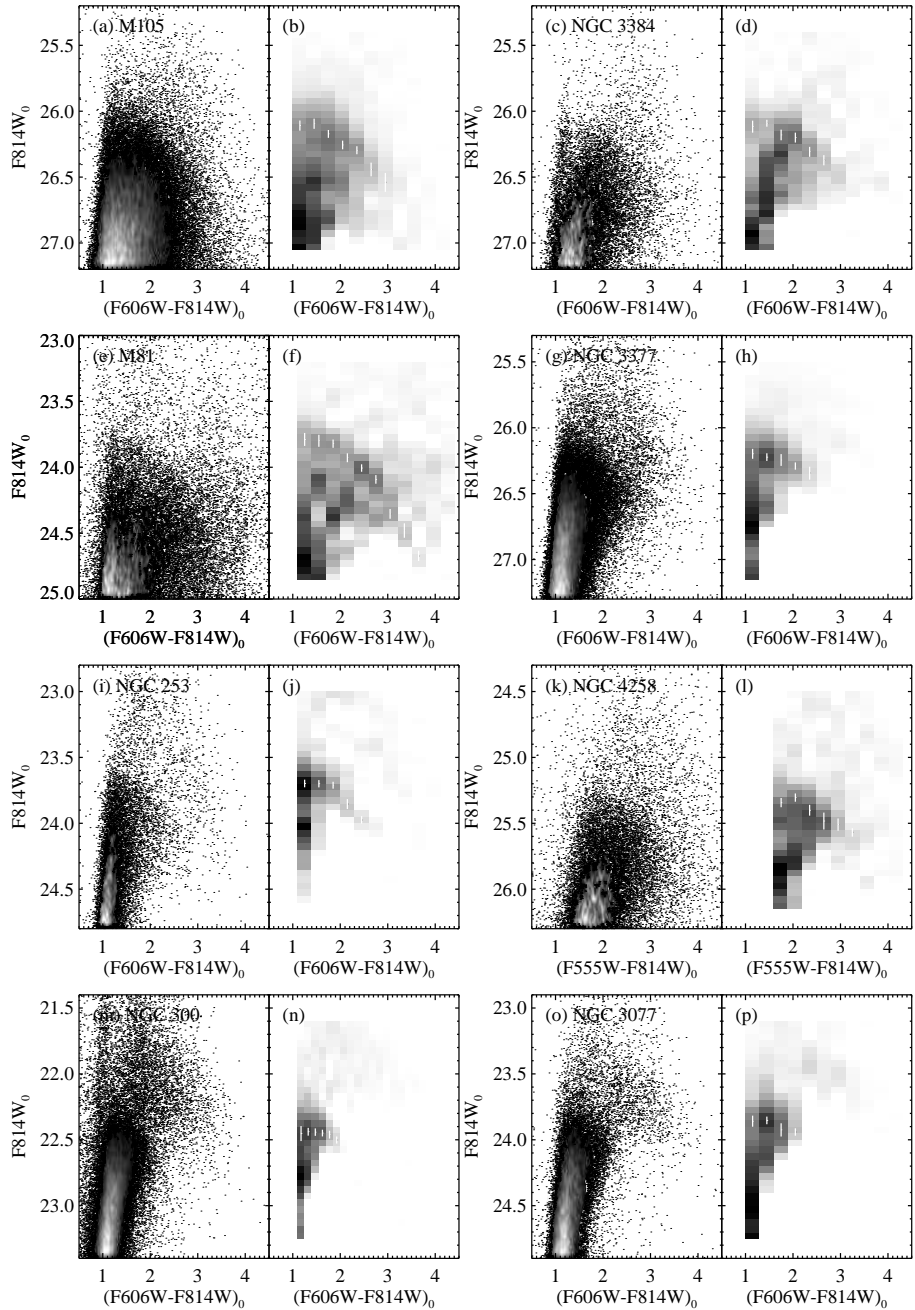} %white.eps}
\caption{%(a, c, e, g, i, k, m, and o) 
(Left windows in each panel) Foreground reddening-corrected CMDs of the resolved stars in eight galaxies used for the color-dependence calibration. 
High density regions in each CMD are displayed as a density map (Hess diagram).
(Right windows in each panel) Maps showing the strength of edge detection responses. 
We divided the stars in each CMD into subgroups based on their color and applied an edge detection algorithm.
Dark and pale regions in each map represent strong and weak edge detection responses, respectively.
Estimated TRGB magnitudes and their errors for the subgroups are marked by white vertical lines. 
Note that $(\rm F555W-F814W)_0$ colors were used for NGC 4258, while $(\rm F606W-F814W)_0$ colors were used for the other galaxies.
}
\label{fig_cmd}
\end{figure*}

In {\color{red}{\bf Figure \ref{fig_cmd} (a, c, e, g, i, k, m, and o)}}, we display CMDs of the resolved stars in the selected regions, marked by hatched regions in {\color{red}{\bf Figure \ref{fig_finder}}}, of each galaxy. 
The central regions of NGC 3384 ($R\leq3\arcmin$), NGC 300 ($R\leq0\farcm6$), and NGC 3077 ($R\leq1\farcm5$) were discarded in the analysis because of serious crowding due to high stellar densities.
The most prominent feature in all CMDs is a distinguishable RGB. The width of the RGBs shows a large variation: from narrow RGBs for low mass galaxies (NGC 300 and NGC 3077) to broad RGBs for high mass galaxies (M105 and M81).
All CMDs show also a small population of  asymptotic giant branch (AGB) stars.
The CMD of the disk fields in NGC 300 shows not only  a strong old RGB but also the presence of red core helium burning stars with much younger age at (F606W--F814W)$_0\approx 1.0$.

%Figure 2

%%%%%%%%%%%%%%%%%%%%%%%%%%%%%%%%%%%%%%%%%%%
%% TABLES 2
%%%%%%%%%%%%%%%%%%%%%%%%%%%%%%%%%
\begin{deluxetable*}{lcc|lcc}
\tabletypesize{\footnotesize}
%\tabletypesize{\scriptsize}
%\tabletypesize{\tiny}
\setlength{\tabcolsep}{0.05in}
%\rotate
\tablecaption{A Summary of the TRGB Magnitudes for the Color Dependence Calibration}
\tablewidth{0pt}
\tablehead{ \colhead{Galaxy} & \colhead{Color range} &  \colhead{F814W$_{0,TRGB}$} & \colhead{Galaxy} & \colhead{Color range} &  \colhead{TRGB}  
%& $F606W-F814W$
}
\startdata
	 	& $(\rm F606W-F814W)_0$ & $F814W_0$& & $(\rm F606W-F814W)_0$& $F814W_0$\\
\multirow{7}{*}{M105} 	& $1.0\sim1.3$ & $26.109\pm0.033$ & \multirow{5}{*}{NGC 253}	& $1.1\sim1.4$ & $23.698\pm0.021$\\
		& $1.3\sim1.6$ & $26.095\pm0.033$ & 	& $1.4\sim1.7$ & $23.700\pm0.019$\\
	 	& $1.6\sim1.9$ & $26.172\pm0.022$ & 	& $1.7\sim2.0$ & $23.711\pm0.017$\\
	 	& $1.9\sim2.2$ & $26.256\pm0.026$ & 	& $2.0\sim2.3$ & $23.854\pm0.032$\\
	 	& $2.2\sim2.5$ & $26.296\pm0.023$ & 	& $2.3\sim2.6$ & $23.974\pm0.020$\\
	 	& $2.5\sim2.8$ & $26.442\pm0.048$ &  \\
	 	& $2.8\sim3.1$ & $26.540\pm0.069$ &  \\
\hline
	 	& $(\rm F606W-F814W)_0$& $F814W_0$&  & $(\rm F555W-F814W)_0$& $F814W_0$\\
\multirow{6}{*}{NGC 3384}& $1.0\sim1.3$ & $26.117\pm0.041$ & \multirow{6}{*}{NGC 4258}& $1.6\sim1.9$ & $25.343\pm0.029$\\
	& $1.3\sim1.6$ & $26.091\pm0.017$ & 	& $1.9\sim2.2$ & $25.305\pm0.024$\\
	& $1.6\sim1.9$ & $26.180\pm0.035$ & 	& $2.2\sim2.5$ & $25.401\pm0.036$\\
	& $1.9\sim2.2$ & $26.200\pm0.032$ & 	& $2.5\sim2.8$ & $25.483\pm0.061$\\
	& $2.2\sim2.5$ & $26.308\pm0.032$ & 	& $2.8\sim3.1$ & $25.510\pm0.042$\\
	& $2.5\sim2.8$ & $26.372\pm0.033$ & 	& $3.1\sim3.4$ & $25.569\pm0.025$\\
\hline
		 	& $(\rm F606W-F814W)_0$& $F814W_0$& 	& $(\rm F606W-F814W)_0$& $F814W_0$\\
\multirow{9}{*}{M81}& $1.1\sim1.4$ & $23.789\pm0.045$ & \multirow{6}{*}{NGC 300}	& $1.10\sim1.25$ & $22.453\pm0.053$\\
		& $1.4\sim1.7$ & $23.799\pm0.041$ & 		& $1.25\sim1.40$ & $22.439\pm0.022$\\
		& $1.7\sim2.0$ & $23.821\pm0.024$ & 		& $1.40\sim1.55$ & $22.442\pm0.020$\\
		& $2.0\sim2.3$ & $23.928\pm0.027$ & 		& $1.55\sim1.70$ & $22.449\pm0.016$\\
		& $2.3\sim2.6$ & $24.006\pm0.020$ & 		& $1.70\sim1.85$ & $22.461\pm0.022$\\
		& $2.6\sim2.9$ & $24.092\pm0.026$ & 		& $1.85\sim2.00$ & $22.502\pm0.018$\\
		& $2.9\sim3.2$ & $24.354\pm0.031$ &  \\
		& $3.2\sim3.5$ & $24.480\pm0.051$ & \\
		& $3.5\sim3.8$ & $24.672\pm0.034$ & \\
\hline
	 	& $(\rm F606W-F814W)_0$& $F814W_0$& & $(\rm F606W-F814W)_0$& $F814W_0$\\
\multirow{5}{*}{NGC 3377}& $1.0\sim1.3$ & $26.198\pm0.032$ & \multirow{4}{*}{NGC 3077}& $1.0\sim1.3$ & $23.859\pm0.037$\\
	& $1.3\sim1.6$ & $26.223\pm0.021$ & 	& $1.3\sim1.6$ & $23.853\pm0.021$\\
	& $1.6\sim1.9$ & $26.245\pm0.040$ & 	& $1.6\sim1.9$ & $23.925\pm0.046$\\
	& $1.9\sim2.2$ & $26.291\pm0.019$ & 	& $1.9\sim2.1$ & $23.941\pm0.020$\\
	& $2.2\sim2.5$ & $26.345\pm0.045$ &\\
%\hline
%\hline
%\hline
\enddata
\label{tab_appendix}
%\tablenotetext{a}{F606W--F814W combination of ACS/WFC}
\end{deluxetable*}

We detected the TRGB by applying the Sobel edge detection algorithm with a kernel of [-1, -2, -1, 0, +1, +2, +1] and a bin size of 0.05 mag.
In order to check the color-magnitude relation of the TRGB, we divided the stars into sub-groups according to their color with a color interval of 0.3 mag. 
In the case of NGC 300, we used a color interval of 0.15 mag because the CMD of this galaxy contains a much larger number of blue RGB stars.
{\color{red}{\bf Figure \ref{fig_cmd} (b, d, f, h, j, l, n, and p)}} shows the TRGB detection results for the target galaxies. Edge detection responses are indicated by grey scale maps: black for strong response (sharp discontinuity in the luminosity function) and white for weak response.
The TRGB magnitudes and corresponding measurement uncertainties are determined through ten thousand simulations of bootstrap resampling. 
Derived values and errors are listed in {\color{red} \bf Table \ref{tab_appendix}} and also indicated by white vertical lines in each grey scale map.
All eight galaxies show clearly color - magnitude relations of the TRGB: the
TRGBs become fainter, as color increases.

\citet{mad09} pointed out that at least $400 \sim 500$ stars in the one mag interval below the TRGB are needed to detect the TRGB within the Poisson noise of 0.1 mag. 
Considering this, we set 100 RGB stars in 0.2 mag interval below the TRGB as a criterion for analysis. We chose a narrower range of magnitude, to use photometry with higher completeness.
Then we counted the number of RGB stars in each subgroup in each galaxy to check whether they have a sufficient number of RGB stars for precise determination of the TRGB or not.
Only the cases satisfying our selection criteria above, as indicated by white vertical lines in {\color{red}{\bf Figure \ref{fig_cmd}}}, were used in the following analysis.

\subsection{Color Transformation between $\rm F555W-F814W$ and $\rm F606W-F814W$}

% Figure 3
\begin{figure}
\centering
\includegraphics[scale=0.88]{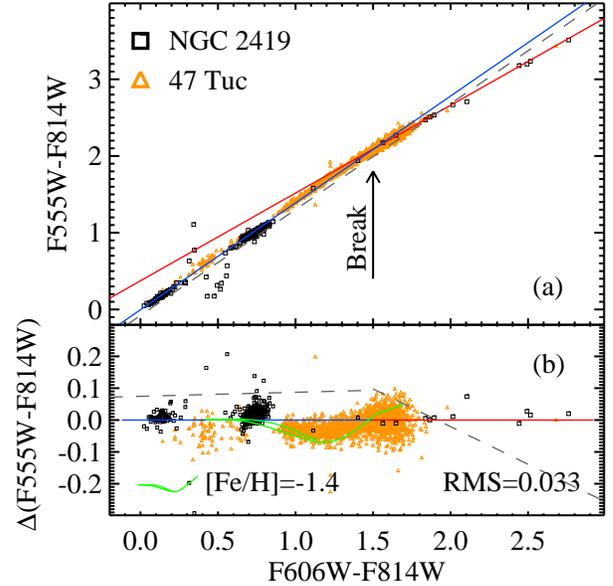} %white.eps}
\caption{
%\color{red}{
(a) Color difference between $\rm F606W-F814W$ and $\rm F555W-F814W$ as a function of $\rm F606W-F814W$ color for the stars in NGC 2419 (open squares) and 47 Tuc (open triangles). 
Blue and red solid lines show double linear fits to the observed colors: the blue stars ($\rm F606W-F814W\leq1.5$, blue line) and the red stars ($\rm F606W-F814W>1.5$, red line).
(b) Residual $\rm F555W-F814W$ color after subtraction of the linear fits for the blue and the red stars.
The curved line indicates 12 Gyr stellar isochrone for a metallicity of [Fe/H]$=-1.4$, which is a mean metallicity of the two globular clusters, provided by the Dartmouth group. 
The gray dashed line represents the color relation expected from the photometric transformations given in \citet{sir05}, which shows a strong difference with respect to the observed stars.}%}
\label{fig_trans}
\end{figure}

% Figure 4
\begin{figure*}
\centering
\includegraphics[scale=1.1]{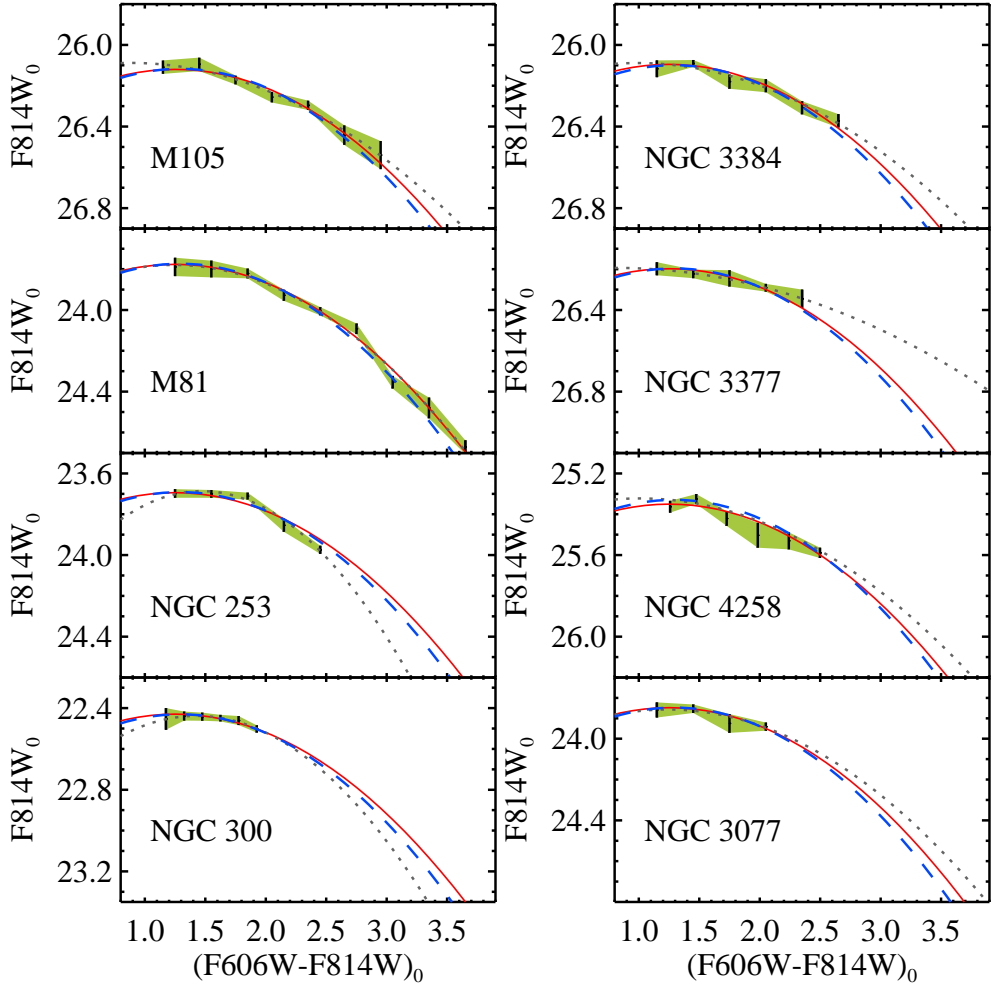} %white.eps}
\caption{
%\color{red}{ 
$\rm F814W_0 - (F606W-F814W)_0$ color-magnitude relations of the TRGB for eight galaxies. Green shaded regions with vertical short lines represent the measured TRGB magnitudes with errors.
Three best fit results with a quadratic equation are overlayed: 
constraining the same curvature parameters ($\alpha$ and $\beta$) for entire color range of all eight galaxies (red solid lines), the same fit results for the red solid lines except for the use of the blue color range (F606W--F814W$\leq 2.5$) only (blue dashed lines), and free curvature parameters for entire color range of each galaxy (grey dotted lines).
}%}
\label{fig_cal1}
\end{figure*}

NGC 4258 is an important galaxy for calibrations of both the color dependence and the zero-point,  %calibrations of the TRGB, 
because it shows a broad range of RGB color and hosts a megamaser, which is useful for geometric distance estimation. NGC 4258 was observed with F555W and $\rm F814W$ filters.
However, the other galaxies used for color dependence calibration were observed with F606W and F814W filters.
Therefore, the color transformation between $\rm F555W-F814W$ and $\rm F606W-F814W$ is needed for a direct comparison.
%{\color{red}
\citet{sir05} provides photometric transformations from ACS   
to Johnson-Cousins systems. 
With these relations, F555W and F814W magnitudes can be converted to $V$ and $I$ magnitudes, respectively.
Similarly, $V$ and $I$ magnitudes can also be converted to F555W and F814W magnitudes, respectively.
This method enables the color transformation between the two ACS colors.
However, applying transformation twice would increase a systematic uncertainty. %}

In Paper III \citep{jan15}, we presented $\rm F555W$, $\rm F606W$, and $\rm F814W$ bands photometry of  two Milky Way globular clusters, NGC 2419 and 47 Tuc.
Using the photometric catalogs of the stars in the two globular clusters obtained from our previous study, we derive a direct color transformation. 
We selected resolved stars in each globular cluster with the following criteria: $R\geq1\arcmin$ and $18.5 < \rm F814W \leq 21.0$ mag for NGC 2419, and $R\geq7\arcmin$ and $19.5 < \rm F814W \leq 22.0$ mag for 47 Tuc (see Figure 4 of \citet{jan15}). 
A comparison of two ACS colors from the selected stars is shown in {\color{red}{\bf Figure \ref{fig_trans}(a)}}. The $\rm F606W - F814W$ color ranges from 0.0 to 2.8 mag. 
A large scatter of NGC 2419 stars at $\rm F606W-F814W\sim0.5$ mag is probably due to the RR Lyrae variables.
The distribution of stars can be fitted well by two linear relations (blue and red solid lines) with a break at $\rm F606W - F814W =1.5$ mag: 
\begin{equation}
(\rm F555W-F814W) = (1.393\pm0.003)(\rm F606W-F814W) - (0.004\pm0.004) \\
\end{equation}
for $\rm F606W-F814W \leq 1.5$ mag, and 
\begin{equation}
(\rm F555W-F814W) = (1.148\pm0.021)(\rm F606W-F814W) + (0.336\pm0.035) 
\end{equation}
for $\rm F606W-F814W > 1.5$ mag.

{\color{red}{\bf Figure \ref{fig_trans}(b)}} displays $\rm F555W-F814W$ color difference between the observed stars and the linear relation of the blue ($\rm F606W-F814W \leq 1.5$ mag) and the red ($\rm F606W-F814W > 1.5$ mag) stars. 
%{\color{red}{
The root mean square of stars over the entire color range (0.0 $<$ F606W--F814W $<$ 2.8) is 0.033 mag. %}}
We overlay a stellar isochrone for 12 Gyr age, and [Fe/H] $= -1.4$, which is a mean metallicity of the two globular clusters, in the Dartmouth models \citep{dot08} as a curved line.
We also overlay a relation from the observational transformations given in \citet{sir05} (dot-dashed line).
Theoretical model agrees very well with the observed stars. 
However, the relation from \citet{sir05} shows a systematic offset of $\sim0.1$ mag.
We use double linear relations derived in this study in the following analysis.

\subsection{Deriving Color - Magnitude relation}

%%%%%%%%%%%%%%%%%%%%%%%%%%%%%%%%%%%%%%%%%%%
%% TABLES 3
%%%%%%%%%%%%%%%%%%%%%%%%%%%%%%%%%
\begin{deluxetable*}{lcrcc}
\tabletypesize{\footnotesize}
%\tabletypesize{\scriptsize}
%\tabletypesize{\tiny}
\setlength{\tabcolsep}{0.05in}
%\rotate
\tablecaption{Approximation of the Color -- Magnitude Relation of the TRGB}
\tablewidth{0pt}
\tablehead{ \colhead{Galaxy} & \colhead{$\alpha$} &  \colhead{$\beta$} & \colhead{$\delta$} & \colhead{RMS} \\
& \multicolumn{3}{c}{($F814W_{TRGB} = \alpha({Color_{TRGB}}^a-1.1)^2 + \beta({Color_{TRGB}}^a-1.1)$} + $\delta$)&
}
\startdata
\multicolumn{5}{l}{Entire color range, the same $\alpha$ and $\beta$ values for all galaxies}\\
M105	  & \multirow{8}{*}{$0.159\pm0.010$}  & \multirow{8}{*}{$-0.047\pm0.020$}  & $26.124\pm0.011$ & 0.026 \\
NGC 3384 & 		   &		      & $26.099\pm0.011$ & 0.028 \\
M81 	 & 		   & 		      & $23.780\pm0.010$ & 0.029 \\
NGC 3377 & 		   & 		      & $26.201\pm0.012$ & 0.024 \\
NGC 253  & 		   & 		      & $23.698\pm0.010$ & 0.035 \\
NGC 4258 & 		   & 		      & $25.357\pm0.011$ & 0.035 \\
NGC 300  & 		   & 		      & $22.433\pm0.009$ & 0.012 \\
NGC 3077 & 		   & 		      & $23.852\pm0.013$ & 0.020 \\
\hline
\multicolumn{5}{l}{Blue RGB stars with F606W--F814W$\leq2.5$, the same $\alpha$ and $\beta$ values for all galaxies} \\
M105	  & \multirow{8}{*}{$0.182\pm0.025$}  & \multirow{8}{*}{$-0.070\pm0.038$}  & $26.124\pm0.011$ & 0.026 \\
NGC 3384 & 		   & 		      & $26.106\pm0.016$ & 0.021 \\
M81 	 & 		   & 		      & $23.780\pm0.016$ & 0.015 \\
NGC 3377 & 		   & 		      & $26.203\pm0.016$ & 0.030 \\
NGC 253  & 		   & 		      & $23.698\pm0.013$ & 0.030 \\
NGC 4258 & 		   & 		      & $25.362\pm0.015$ & 0.037 \\
NGC 300  & 		   & 		      & $22.437\pm0.012$ & 0.010 \\
NGC 3077 & 		   & 		      & $23.854\pm0.017$ & 0.022 \\
\hline
\multicolumn{5}{l}{Entire color range, different $\alpha$ and $\beta$ values for each galaxy }\\
M105     & $0.106\pm0.050$ & $ 0.047\pm0.085$ & $26.093\pm0.035$ & 0.027 \\
NGC 3384 & $0.107\pm0.052$ & $ 0.024\pm0.088$ & $26.090\pm0.039$ & 0.029 \\
M81 	 & $0.157\pm0.050$ & $-0.044\pm0.070$ & $23.782\pm0.046$ & 0.031 \\
NGC 3377 & $0.063\pm0.166$ & $ 0.037\pm0.235$ & $26.199\pm0.060$ & 0.026 \\
NGC 253  & $0.307\pm0.084$ & $-0.229\pm0.132$ & $23.729\pm0.042$ & 0.034 \\
NGC 4258 & $0.116\pm0.074$ & $ 0.020\pm0.134$ & $25.338\pm0.059$ & 0.035 \\
NGC 300  & $0.249\pm0.252$ & $-0.164\pm0.262$ & $22.465\pm0.061$ & 0.012 \\
NGC 3077 & $0.135\pm0.150$ & $-0.037\pm0.155$ & $23.859\pm0.042$ & 0.020 \\
\multicolumn{1}{l}{Weighted mean} & $0.139\pm0.025$ & $-0.023\pm0.039$ \\
\enddata
\label{tab_fits}
\tablenotetext{a}{F606W--F814W filter combination of ACS/WFC}
\end{deluxetable*}

In {\color{red}{\bf Figure \ref{fig_cal1}}}, we plotted the $\rm (F606W-F814W)_0$ color and F814W$_0$ magnitude relations of the TRGB stars in eight galaxies. The $\rm (F555W-F814W)_0$ color of NGC 4258 was converted to $\rm (F606W-F814W)_0$ color using the transformation described in Section 2.3. The TRGB magnitudes show a clear non-linear relation as a function of color.
We use an equation below, keeping quadratic terms for the approximation of the TRGB: 
\begin{equation}
\rm F814W_{TRGB} = \alpha(Color_{TRGB} - \gamma)^2 + \beta(Color_{TRGB} - \gamma) + \delta
\end{equation}

\noindent where $\gamma$ means a fiducial color at [Fe/H] $=-1.6$. 
In the case of $\rm F606W-F814W$ combination, the value of $\gamma$ corresponds to 1.1. 
Coefficients $\alpha$ and $\beta$ determine the shape of the relation, and $\delta$ reflects apparent magnitudes of the TRGB in each galaxy. 
Thus, values of $\alpha$ and $\beta$ are the same in all eight galaxies, but that of $\delta$ depends on galaxies. We carried out a multi-parameter fitting for entire color range of the TRGB bins using the IDL/mpfitexpr to find the values of three coefficients and listed results in {\color{red}{\bf Table \ref{tab_fits}}}.
Derived values are $\alpha=0.159\pm0.010$, $\beta=-0.047\pm0.020$, and $\delta= 22.4\sim26.2$, as plotted by solid lines in {\color{red}{\bf Figure \ref{fig_cal1}}}.

% Figure 5
\begin{figure}
\centering
\includegraphics[scale=0.9]{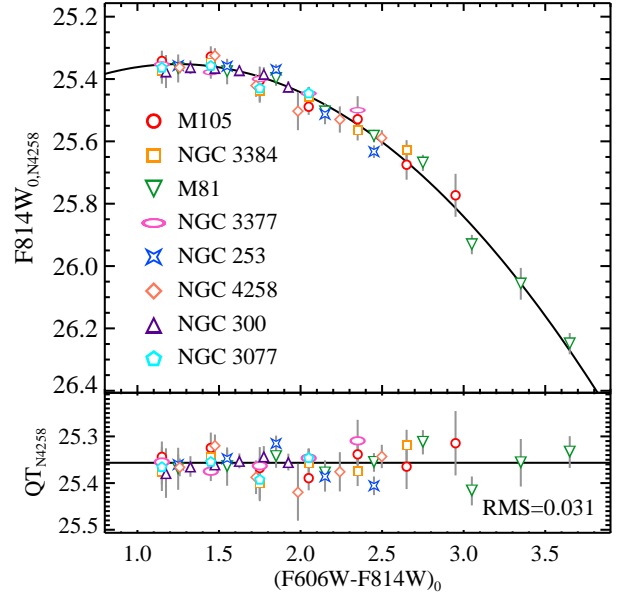} %white.eps}
\caption{(Top) Combined $\rm F814W_0$ magnitudes of the TRGB for eight galaxies as a function of $(\rm F606W-F814W)_0$ color. $\rm F814W_0$ magnitudes of each galaxy were shifted along the vertical direction, to be consistent with the result from NGC 4258. The solid line denotes the best fit result.
(Bottom) Residuals between the best fit result and the data. Note that the RMS value for fitting is as  small as 0.027 mag.
}
\label{fig_cal2}
\end{figure}

%{\color{red}{
We investigated the variation of $\alpha$ and $\beta$ values depending on the selection of the TRGB color and the galaxies.
When we use the TRGB color bins bluer than $(F606W-F814W)_0=2.5$ in the fit, we get $\alpha=0.182\pm0.025$ and $\beta=-0.070\pm0.038$ (shown as dashed lines in {\color{red}{\bf Figure \ref{fig_cal1}}}), 
which agree well with those from the entire color bins.
If we try fits for individual galaxies with the entire color bins, we obtain $\alpha=0.063$ to $0.307$, $\beta= -0.229$ to $0.047$ (as shown by dotted lines in the figure). 
Weighted mean of the eight $\alpha$ and $\beta$ values are $\alpha=0.139\pm0.025$ and $\beta=-0.023\pm0.039$, which are similar to the results from the same curvature parameters for all eight galaxies. 
All the $\alpha$ and $\beta$ values from individual galaxies agree well within 1$\sigma$ level, with the weighted mean values, except for the values from NGC 253. 
NGC 253 gives $\alpha=0.307\pm0.084$ and $\beta=-0.229\pm0.132$, which show 2.0$\sigma$ and 1.6$\sigma$ differences, marginally agreeing with the weighted mean values.
In the following analysis, we use $\alpha$ ($=0.159\pm0.010$) and $\beta$ ($=-0.047\pm0.020$) values from the fit of the entire color range of the TRGB bins with the same $\alpha$ and $\beta$ values of all eight galaxies.
%}}

We shifted the TRGB magnitudes and corresponding fit results of each galaxy along the vertical direction, to be consistent with the result from NGC 4258, and plotted them in {\color{red}{\bf Figure \ref{fig_cal2}(a)}}. All eight galaxies show reasonable agreements.
The difference between the data points and the best fit result is shown in {\color{red}{\bf Figure \ref{fig_cal2}(b)}}. 
The root mean square of data is estimated to be 0.031 mag. It is similar to the mean measurement uncertainty of the TRGB.

We computed the correction parameters $\alpha$, $\beta$, and $\gamma$ for several different filter systems as well. 
The values for $\rm F555W-F814W$ combination in ACS/WFC were determined using the color transformation between $\rm F555W-F814W$ and $\rm F606W-F814W$ shown in Section 2.3.
The transformation was applied to seven galaxies that were observed with the F606W and F814W filters, excluding NGC 4258.
Then we carried out multi-parameter fitting as done for $\rm F606W-F814W$ combination and obtained the values. In the fitting, we fixed $\gamma=1.6$, which corresponds to [Fe/H] $=-1.6$.
%{\color{red}{
Conversion relations described in \citet{sir05} were used to obtain the parameters in three photometric systems: $\rm F555W-F814W$ and $\rm F606W-F814W$ combinations in WFPC2, and $V-I$ combination in the Johnson-Cousins system. 
Parameters for  $\rm F555W-F814W$ and  $\rm F606W-F814W$ combinations in WFC3/UVIS were derived by applying the photometric transformation between ACS/WFC to WFC3/UVIS described in \citet{jan15}. Transformation uncertainty, $0.01\sim0.02$ mag were added in each calculation. Derived values are summarized in {\color{red} \bf{Table \ref{tab_cal}}}.
%}}

The color magnitude relation of the TRGB given in {\color{red}\bf Figure 5} shows two interesting properties. 
First, at the blue color range of $F606W-F814W\leq 1.5$ (corresponding to $F555W-F814W\leq 2.1$ and $V-I\leq1.9$), color magnitude relation of the TRGB is nearly flat in F814W$_0$ -- (F606W--F814W)$_0$ CMD, which means that additional color (metallicity) correction is not required in the TRGB measurement.
If we use these blue RGB stars only in the TRGB measurement, distance modulus can be derived directly by adding the zero-point of the TRGB in F814W:
\begin{equation}
(m-M)_0 = F814W_{0,TRGB} + M_{F814W,TRGB}.
\end{equation}
Hereafter, we call this the blue $I$ calibration.

Second, the red color range with $F606W-F814W>1.5$ shows a color magnitude dependence, so that additional color correction should be applied for an accurate calibration of the TRGB.

We introduce the $QT$ magnitude, a quadratic form of the TRGB magnitude corrected for the color dependence of the F814W (or $I$) magnitude TRGB.
The $QT$ is expressed by
\begin{equation}
QT =  \rm F814W_{0} - \alpha(Color-\gamma)^2 - \beta(Color-\gamma)
\end{equation}

\noindent where $\alpha=0.159\pm0.010$, $\beta=-0.047\pm0.020$, and $\gamma= 1.1$ in the $\rm F606W-F814W$ filter combination in ACS/WFC.
Distance modulus is obtained as follows:

\begin{equation}
(m-M)_0 = QT_{RGB} + M_{QT,TRGB}
\end{equation}
\noindent where $M_{QT,TRGB}$ is an absolute magnitude of the TRGB in the $QT$ system. 
This TRGB calibration is called hereafter the $QT$ calibration.
Values of $M_{F814W,TRGB}$ and $M_{QT,TRGB}$ will be addressed in the next Section.

\section{Zero-point Calibration}

\subsection{NGC 4258 as a Distance Anchor}

NGC 4258 is known to be a useful distance anchor because it hosts a water megamaser, which can be used as a powerful and precise geometric distance indicator. 
With monitoring observations of the nuclear water megamaser sources in NGC 4258,  several geometric distance estimates were presented: \citet{gre95, miy95, her99, hum05, arg07, hum08, hum13, rie16}. 
\citet{hum13} reported a distance value of $d=7.60\pm0.17_r\pm0.15_s$ Mpc ($(m-M)_0=29.404\pm0.049_r\pm0.043_s$), where errors with $r$ and $s$ represent random errors and systematic errors, respectively.
It is slightly larger than the old value given by \citet{her99} ($d=7.2 \pm 0.2_r \pm 0.5_s$ Mpc), which is similar to the value used in \citet{rie11,rie12} ($d=7.28\pm0.22$ Mpc). 
Most recently, \citet{rie16} provided 
a similar value with smaller systematic uncertainty, d=$7.54\pm0.17_r\pm0.10_s$ Mpc ($(m-M)_0=29.387\pm0.049_r\pm0.029_s$).
%{\color{red}{
They performed a significantly larger MCMC analysis than \citet{hum13}, reducing the systematic uncertainty by a factor of two. %}}
We adopted the most recent distance estimate given by \citet{rie16} for the zero-point calibration of the TRGB in this study.

\citet{mag08} reported a TRGB distance to NGC 4258 based on  photometry of the same HST images as used in this study. They carried out PSF photometry using two software tools, DOLPHOT \citep{dol00} and DAOPHOT, and derived a TRGB magnitude, $T_{RGB}=25.24\pm0.04$ from DOLPHOT photometry. It is noted that they found $\sim0.04$ mag difference in the TRGB magnitude between the two data reduction methods.

%%%%%%%%%%%%
% Fig 6
%%%%%%%%%%%%
\begin{figure*}
\centering
\includegraphics[scale=0.9]{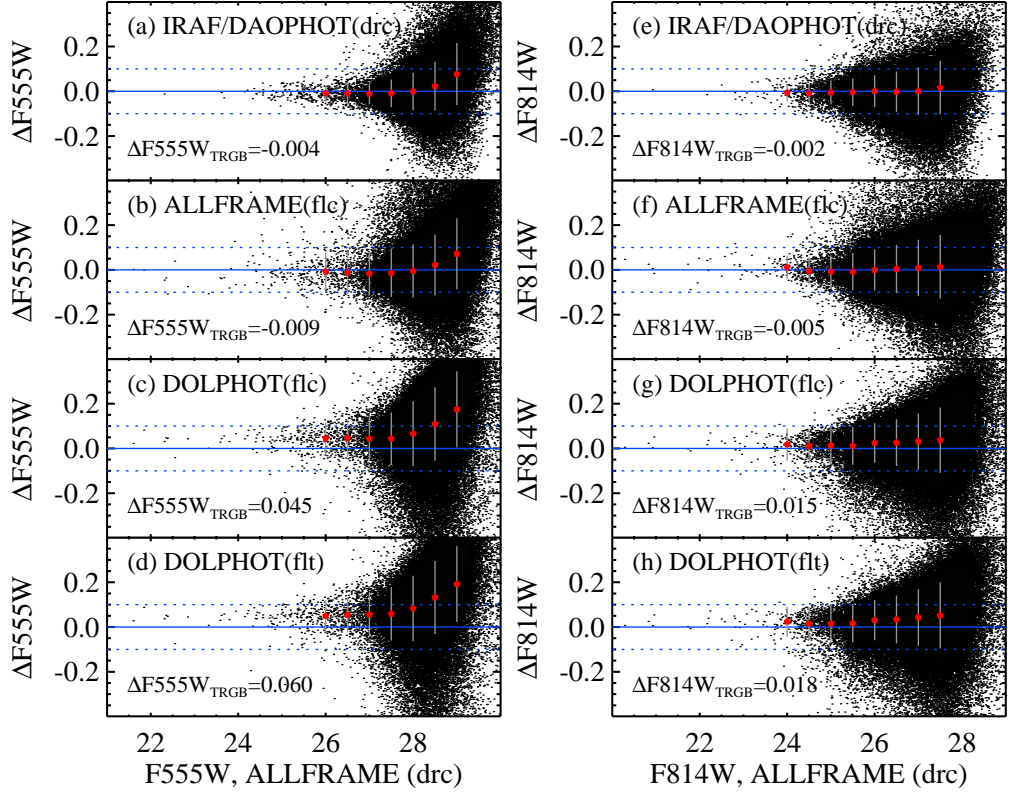} %white.eps}
\caption{
(a-d) Comparisons of PSF magnitudes in $\rm F555W$ band between ALLFRAME on drc reduction and IRAF/DAOPHOT on drc (a), ALLFRAME on flc (b), DOLPHOT on flc (c), and DOLPHOT on flt (d) reductions of NGC 4258. 
Red dots and error bars represent the mean magnitude offset and its standard deviation, respectively. Mean magnitude offsets at the TRGB level ($\rm F555W\sim28.0$) are labeled in each panel. (e-h) Same as (a-d), except for F814W band. Note the mean magnitude offsets at the TRGB level ($\rm F814W \sim25.3$) are very small, $\Delta \rm F814W \lesssim 0.02$ mag.
}
\label{fig_m106_compare}
\end{figure*}

%%%%%%%%%%%%
% Fig 7
%%%%%%%%%%%%
\begin{figure*}
\centering
\includegraphics[scale=1.0]{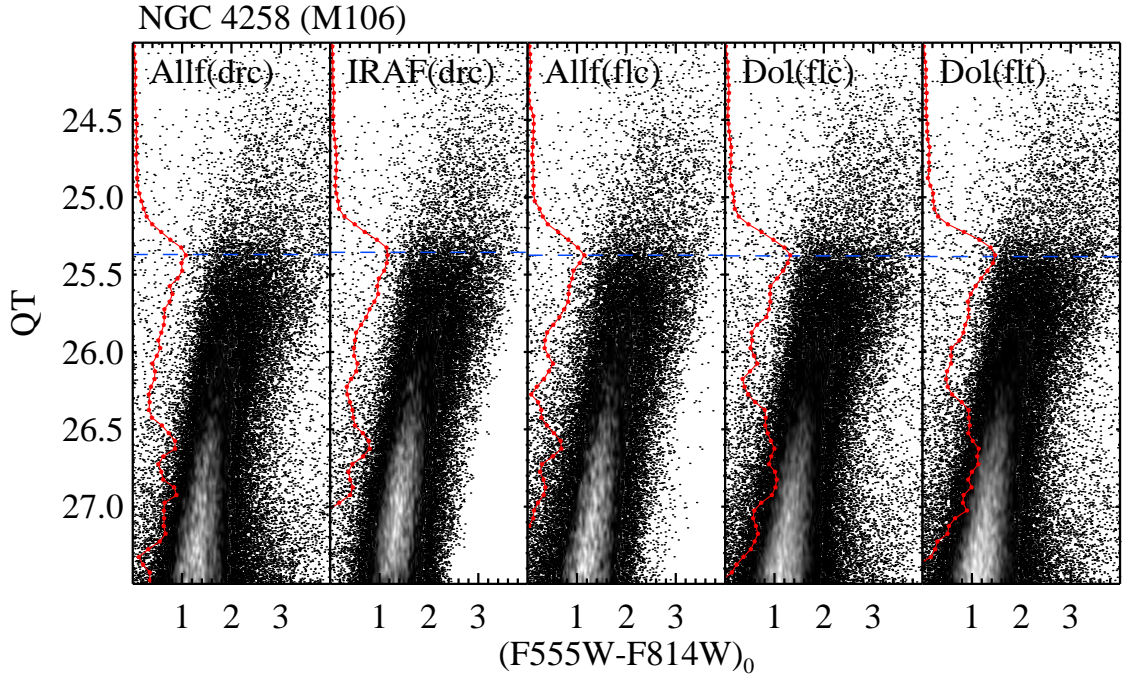} %white.eps}
\caption{$QT-(\rm F555W-F814W)_0$ CMDs of NGC 4258 from five different reduction methods : ALLFRAME on drc, IRAF/DAOPHOT on drc, ALLFRAME on flc, DOLPHOT on flc, and DOLPHOT on flt (from left to right). Edge detection responses are shown by the solid lines.
Note that the estimated TRGB magnitudes (dashed lines) agree very well.
}
\label{fig_cmd_m106}
\end{figure*}

We obtained a value for the TRGB magnitude in the $QT$ system for NGC 4258 from  photometry of ACS image data that we used for the color dependence calibration (Section 2). 
We also performed a test similar to the \citet{mag08}'s. 
We performed point source photometry based on five different techniques using different softwares and different image types: 
1) DAOPHOT/ALLFRAME run on charge transfer efficiency (CTE) corrected drizzled images (indicated by \_drc.fits), 
2) IRAF/DAOPHOT run on the same \_drc.fits images, 
3) DAOPHOT/ALLFRAME run on CTE corrected and flat fielded images (indicated by \_flc.fits), 
4) DOLPHOT run on the same \_flc.fits images, and
5) DOLPHOT run on CTE uncorrected and flat fielded images (indicated by \_flt.fits).
We used PSF images we derived using isolated bright stars (for DAOPHOT/ALLFRAME and IRAF/DAOPHOT reductions on drizzled images) or provided by the Tiny Tim (for DOLPHOT reductions).
In the case of DAOPHOT/ALLFRAME photometry on individual \_flc.fits images, we used PSF images constructed from the ACS images of 47 Tuc (Proposal IDs = 10737 for F555W and 10101 for F814W filters), because the number of bright stars in a single \_flc.fits image of NGC 4258 is only a few.
A single pass sequence of DAOFIND-PHOT-ALLSTAR/ALLFRAME routines was applied for the IRAF/DAOPHOT and ALLFRAME reductions.
DOLPHOT reductions were done using the parameter set recommended by the DOLPHOT/ACS User's guide (version 2.0). 
We set the ACSuseCTE = 0 (CTE uncorrection mode) for \_flc.fits and 1 (CTE correction mode) for \_flt.fits in DOLPHOT reductions.

{\color{red}{\bf Figure \ref{fig_m106_compare}}} displays the magnitude differences between the results of the ALLFRAME (drc) reduction and those of other reductions. 
Comparisons of F555W band magnitudes (left panels) show small offsets.
Mean offsets for the bright stars with F555W $\lesssim27.0$ are smaller than 0.05 mag.
At the TRGB level, F555W $\sim27.8$ mag, mean offsets are estimated to be --0.004, --0.009, 0.045, and 0.060 mag for the IRAF/DAOPHOT (drc), ALLFRAME (flc), DOLPHOT (flc), and DOLPHOT (flt), respectively. 
Thus, three DAOPHOT reductions (ALLFRAME (drc), IRAF/DAOPHOT (drc), and ALLFRAME (flc)) agree well within 0.01 mag, but DOLPHOT reductions show $\sim0.05$ mag differences.
%, which are still small enough.
Faint stars with F555W $\gtrsim28.0$ mag yield larger offsets, which increase as a function of magnitude.
The origin of these magnitude differences is unclear.
Although non-negligible offsets are detected at the faint levels ($F555W \gtrsim28.0$), most of the RGB stars we used in the TRGB detection of NGC 4258 are brighter than F555W $\sim28.0$, where mean offsets are small enough. 
Moreover, $F555W$ band magnitudes are only used in the estimation of the TRGB color, which affects little in the $QT$ magnitude estimation.
Comparisons of F814W band magnitudes show much better agreements.
All five reductions agree well within $0.02$ mag from the bright to the faint magnitude range ($21.0\lesssim \rm F814W \lesssim 27.5$ mag).

In {\color{red}{\bf Figure \ref{fig_cmd_m106}}} we compare $QT -(\rm F555W -F814W)_0$ CMDs from five different reductions. All five CMDs show strong edge detection responses at $QT \sim 27.4$, which represents clearly a TRGB. 
ALLFRAME reductions (drc and flc) show sharper responses at $QT \sim 27.4$ than that of IRAF/DAOPHOT (drc). 
Precise TRGB magnitudes and uncertainties are obtained using the bootstrap method as done in \citet{jan15}:  
$QT_{RGB} = 25.370\pm0.023, 25.356\pm0.053, 25.375\pm0.021, 25.379\pm0.018, 25.283\pm0.019$ mag for ALLFRAME(drc), IRAF/DAOPHOT (drc), ALLFRAME (flc), DOLPHOT (flc), and DOLPHOT(flt), respectively. 
These estimations yield  a mean of $QT_{RGB}=25.373$ mag, median of $QT_{RGB}=25.375$ mag, and standard deviation of 0.010 mag, which is very small. Thus our TRGB estimation is robust. Software-dependent uncertainties in the TRGB detection are not significant. 
We use the TRGB magnitude from the ALLFRAME (drc) reduction, $QT_{RGB} = 25.370\pm0.023$ in the following analysis.

\begin{figure}
\centering
\includegraphics[scale=0.9]{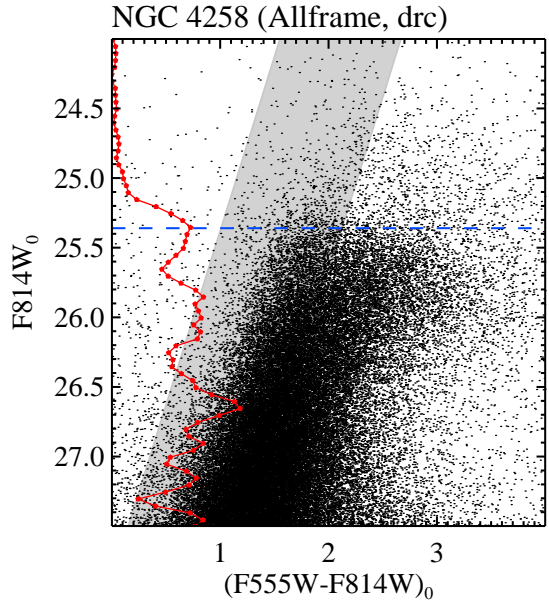} %white.eps}
\caption{F814W$_0$ --(F555W-F814W)$_0$ CMD of NGC 4258 from the ALLFRAME (drc) reduction. 
An edge detection response for resolved stars in the shaded region is shown by a solid line.
The TRGB is estimated to be at F814W$_0 = 25.36$ mag (dashed line).}
\label{fig_cmd_m106i}
\end{figure}

%{\color{red}{
We also determine a color dependence uncorrected TRGB magnitude, F814W$_{0,TRGB}$.
{\color{red}\bf Figure \ref{fig_cmd_m106i}} shows F814W$_0$ -- (F555W--F814W)$_0$ CMD of NGC 4258 from the ALLFRAME (drc) reduction.
We selected the blue RGB stars in the shaded region where color dependence of the TRGB is less severe ($(\rm F555W-F814W)_{0} \leq 2.1$) and plotted their edge detection response by a solid line. % in Figure \ref{fig_cmd_m106i}.
The edge detection response is relatively noisy compared to that from the $QT$ CMD, but still shows
a major peak at $\rm F814W_0=25.357\pm0.031$ mag, which is the TRGB.%}}

%%%%%%%%%%%%
% Fig 8
%%%%%%%%%%%%

From the TRGB magnitudes derived in this study ($QT = 25.370\pm0.023$ mag and $\rm F814W_0=25.357\pm0.031$ mag) and the distance modulus to NGC 4258 given by \citet{rie16}, we determine the zero-points of the TRGB: $M_{\rm {QT},TRGB} = -4.023\pm0.073$ mag for the $QT$ calibration and $M_{\rm {F814W},TRGB} = -4.036\pm0.074$ mag for the blue $I$ calibration.
%{\color{red}{
Their uncertainties, 0.073 mag and 0.074 mag are obtained from the quadratic sums of individual uncertainties as summarized in {\color{red}{\bf Table \ref{tab_error}}}:
TRGB detection errors (0.023 mag and 0.031 mag), 
an extinction correction error  (0.003 mag, 10\% of F814W band extinction), 
a photometric zero-point error (0.020 mag), 
a color transformation error (0.005 mag), 
an F814W to $QT$ transformation error (0.014 mag), 
and an NGC 4258 distance  error (0.057 mag). 
When we use the $QT$ calibration in the standard $VI$ system, a photometric transformation error (0.03 mag) is added. 
Thus systematic uncertainties are dominated by the NGC 4258 distance error.%}}

%%%%%%%%%%%%%%%%%%%%%%%%%%%%%%%%%%%%%%%%%%%
%% TABLES 4
%%%%%%%%%%%%%%%%%%%%%%%%%%%%%%%%%
\begin{deluxetable}{lcccc}
\tabletypesize{\small}
\tabletypesize{\footnotesize}
%\tabletypesize{\scriptsize}
%\tabletypesize{\tiny}
\setlength{\tabcolsep}{0.05in}
%\rotate
\tablecaption{Error Budget for the $QT$ Calibration of the TRGB in the Standard $VI$ system}
\tablewidth{0pt}
\tablehead{ \colhead{Error source} &  \multicolumn{2}{c}{$QT$ calibration} & \multicolumn{2}{c}{Blue $I$ calibration} \\% \colhead{$\omega$ Centauri}  
\colhead{ } &  \colhead{NGC 4258} & \colhead{$LMC$} & \colhead{NGC 4258} & \colhead{$LMC$}
%& [mag] & [mag] &
}
 \startdata
TRGB detection 	& 0.023 & 0.020  & 0.031 & 0.042  \\
Extinction		& 0.003 & 0.07   & 0.003 & 0.07 \\ %& 0.04	&  \\
Photometric zero-point& 0.02 & 0.03 & 0.02 & 0.03\\ %& ...   &  \\
%$I$ to F814W transformation & ... & 0.03 \\
Color transformation$^a$ & 0.005 & ... & ... & ...\\
%F555W--F814W to F606W-F814W transformation &  & \\
F814W to $I$ transformation & 0.03 & ... & 0.03 & ...\\
$QT$ transformation & 0.014 & 0.014 & ... & ...\\
%$QT$ transformation & 0.014$^b$ & 0.005$^c$ \\
%$I$ to $QT$ transformation & ... & 0.005 \\

Intermediate-age population & ...	& 0.02 & ... & 0.02 \\
Distance anchor	& 0.057 & 0.049 & 0.057 & 0.049 \\ %& 0.11  &  \\
\hline
Total			& 0.073 & 0.096 & 0.074 & 0.102  \\
%\multicolumn{2}{c}{Weighted }	 & 0.065 & 0.09 &  \\
\enddata
\label{tab_error}
\tablenotetext{a}{F555W--F814W to F606W--F814W transformation}
%\tablenotetext{b}{F814W to $QT$ transformation}
%\tablenotetext{c}{$I$ to $QT$ transformation}
\end{deluxetable}

% 25.272, 25.274, 25.289, 25.292, 25.290
% mean 25.283
% median 25.289
% stdev 0.010

\subsection{The Large Magellanic Cloud as a Distance Anchor}

%%%%%%%%%%%%
% Fig 9
%%%%%%%%%%%%

\begin{figure}
\centering
\includegraphics[scale=0.9]{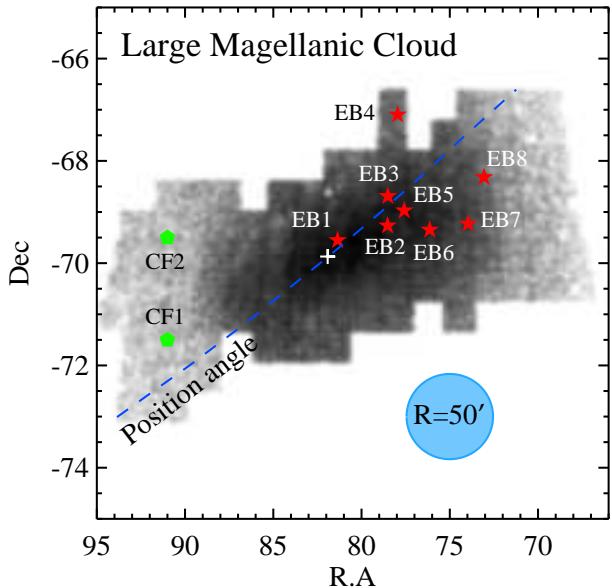} %white.eps}
\caption{Identification of eight eclipsing binary stars (starlets) used in \citet{pie13}, overlayed on a number density map of RGB stars in the LMC. Two control fields used in this study are marked by green pentagons. We investigated the TRGB magnitude of the LMC using the stars at $R\leq50\arcmin$ region of each eclipsing binary star and each control field.
}
\label{fig_finding_lmc}
\end{figure}

%%%%%%%%%%%%
% Fig 10
%%%%%%%%%%%%
\begin{figure}
\centering
\includegraphics[scale=0.7]{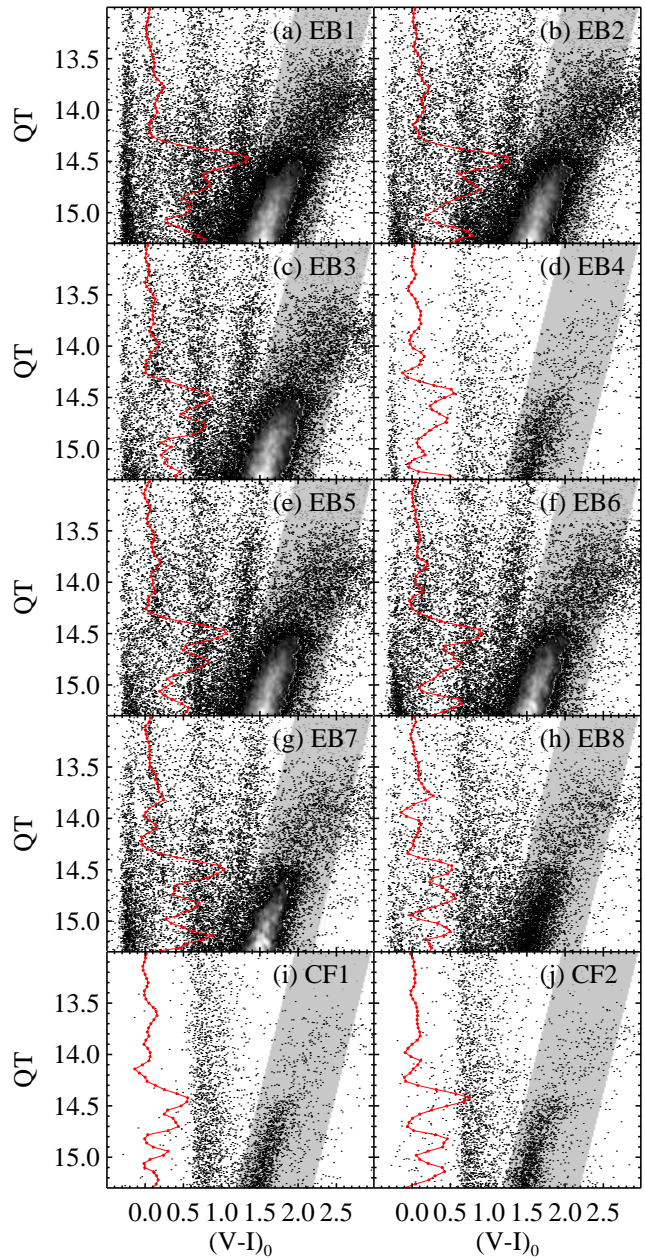} %white.eps}
\caption{$QT - (V-I)_0$ CMDs of resolved stars in $R\leq50\arcmin$ regions of eclipsing binary stars (a-h) and control fields (i-j). Shaded regions indicate the area we used to sample the RGB stars for the TRGB analysis.
Edge detection responses are shown by the solid lines.
}
\label{fig_cmd_lmc}
\end{figure}

The LMC has been often used as a distance anchor for extragalactic distance indicators \citep{fre10,deg14}.
Its distance has long been controversial from $(m-M)_0=18.1$ to $18.8$, and it appeared to settle down with a value of $(m-M)_0\sim18.50\pm0.10$, after the results of the Hubble Key Project \citep{fre01} were published.
However, it has been suspected that the tight correlation of the LMC distance estimates might have been affected by the publication bias or the bandwagon effect \citep{sch08,deg14}.
This problem can be resolved with more accurate distance estimates for the LMC in the future.

Recently \citet{fre12} presented an updated distance estimate, $(m-M)_0=18.477\pm 0.011_r\pm0.033_s$, accurate to 1.7\%, from the analysis of Spitzer $3.6\mu m$ photometry for 10 Cepheids in the Milky Way and 80 Cepheids in the LMC.
On the other hand, \citet{pie13} analyzed eight %carefully selected
 eclipsing binary stars using the OGLE photometry, and yielded a distance value with a total uncertainty of 2.2\%: $(m-M)_0=18.494\pm 0.008_r\pm0.048_s$ ($d=49.97\pm0.19_r\pm0.11_s$ kpc).
These two recent estimates agree very well.
We adopted the distance estimate by \citet{pie13} for the zero-point calibration of the TRGB, because we
use the OGLE photometry of the LMC for the analysis of the TRGB in this study as follows.

We measured the TRGB magnitude of the LMC using the $VI$ photometric catalog  from the OGLE-III shallow survey \citep{ula12}.
This catalog was made using the same telescope and the same CCD camera as used for the analysis of \citet{pie13}.
It offers benefits that any instrument-dependent uncertainties, which are hard to measure, can be ruled out in the zero point calibration of the TRGB.
{\color{red}{\bf Figure \ref{fig_finding_lmc}}} shows identifications of eight eclipsing binary stars used in \citet{pie13} (starlets) overlayed on a spatial number density map of bright RGB stars ($1.2 < V-I \leq 2.0$ and $14.4<  I \leq 16.0$) in the OGLE-III catalog. 
All eight binary stars are located on the western side from the LMC center (white cross).
It is known that the distance measurement can be affected by the geometry and the depth of the LMC.
For this reason, \citet{pie13} applied additional geometry correction to their distance estimation. However, they concluded that the effect of this correction to the distance measurement is negligible.
For the analysis of the TRGB for the LMC we selected  eight fields located around the eclipsing binary stars used in \citet{pie13}, and used the stars at $R\leq50\arcmin$ region around each eclipsing binary star.
We investigated also two additional fields, control field 1 and 2 (pentagons), covering the eastern side of the LMC as a reference.

%%%%%%%%%%%%%%%%%%%%%%%%%%%%%%%%%%%%%%%%%%%
%% TABLE 5
%%%%%%%%%%%%%%%%%%%%%%%%%%%%%%%%%
\begin{deluxetable*}{llccccccc}
\tabletypesize{\scriptsize}
%\tabletypesize{\tiny}
\setlength{\tabcolsep}{0.05in}
%\rotate
\tablecaption{A Summary of the TRGB Magnitudes of the LMC}
\tablewidth{0pt}
\tablehead{ \colhead{Field} & \colhead{R.A.} & \colhead{Decl} &  \multicolumn{2}{c}{$QT^a$ magnitude} &  \multicolumn{2}{c}{$I^b$ magnitude}\\
& & & (Measured) & (Distortion corrected) & (Measured) & (Distortion corrected)
}
\startdata
EB1 	& 05 25 25.550 & -69 33 04.39 	& $14.485\pm0.010$ & $14.489\pm0.010$ & $14.521\pm0.035$ & $14.525\pm0.035$\\
EB2 	& 05 14 05.952 & -69 15 56.83 	& $14.486\pm0.011$ & $14.482\pm0.011$ & $14.522\pm0.017$ & $14.518\pm0.017$\\
EB3 	& 05 14 01.908 & -68 41 18.41 	& $14.492\pm0.012$ & $14.496\pm0.012$ & $14.527\pm0.025$ & $14.531\pm0.025$\\
EB4 	& 05 11 49.458 & -67 05 45.19 	& $14.459\pm0.012$ & $14.485\pm0.012$ & $14.475\pm0.013$ & $14.501\pm0.013$\\
EB5 	& 05 10 19.650 & -68 58 12.00 	& $14.502\pm0.014$ & $14.498\pm0.014$ & $14.480\pm0.023$ & $14.476\pm0.023$\\
EB6 	& 05 04 32.882 & -69 20 50.99 	& $14.505\pm0.015$ & $14.489\pm0.015$ & $14.580\pm0.015$ & $14.564\pm0.015$\\
EB7 	& 04 55 51.491 & -69 13 47.99 	& $14.511\pm0.016$ & $14.487\pm0.016$ & $14.569\pm0.016$ & $14.565\pm0.016$\\
EB8 	& 04 52 15.280 & -68 19 10.30 	& $14.524\pm0.016$ & $14.509\pm0.016$ & $14.472\pm0.027$ & $14.457\pm0.027$\\
CF1 	& 06 04 00.000 & -71 30 00.00 	& $14.470\pm0.030$ & $14.485\pm0.030$ & $14.468\pm0.025$ & $14.483\pm0.025$\\
CF2 	& 06 04 00.000 & -69 30 00.00	& $14.425\pm0.015$ & $14.472\pm0.015$ & $14.472\pm0.021$ & $14.519\pm0.021$\\
\hline

%\multicolumn{3}{l}{Weighted mean of EB1-8 and CF1-2} 			& $14.464\pm0.006$ & $14.466\pm0.006$ \\
\multicolumn{3}{l}{Weighted mean of EB1-8 and CF1-2} 			& $14.485\pm0.004$ $(0.030)^c$ & $14.489\pm0.004$ $(0.011)$ & $14.515\pm0.006$ (0.042) & $14.522\pm0.006$ (0.036)\\
%\multicolumn{3}{l}{Weighted mean of CF1-2} 			& $14.470\pm0.014$ & $14.497\pm0.014$ \\
\multicolumn{3}{l}{Weighted mean of CF1-2} 			& $14.432\pm0.013$ $(0.025)$ & $14.475\pm0.013$ $(0.009)$ & $14.470\pm0.016$ (0.003) & $14.504\pm0.016$ (0.026)\\
%\multicolumn{3}{l}{Standard deviation of EB1-8 and CF1-2} 			& 0.028 & 0.028 \\
%\multicolumn{3}{l}{Weighted mean of EB1-8} 			& $14.463\pm0.007$ & $14.458\pm0.007$ \\
\multicolumn{3}{l}{Weighted mean of EB1-8} 			& $14.491\pm0.004$ $(0.020)$ & $14.490\pm0.004$ $(0.009)$ & $14.523\pm0.007$ (0.042) & $14.524\pm0.007$ (0.039)\\
%\multicolumn{3}{l}{$M_{I,TRGB}$} 	& -4.031 & -4.035 \\
\hline
\enddata
%\tablenotetext{a}{A }
\tablenotetext{a}{A wide color range of the TRGB with $1.5 \leq (V-I)_{0,TRGB}\leq 2.5$ was used.}
%\tablenotetext{a}{TRGB stars with $1.5 \leq (V-I)_{0,TRGB}\leq 2.5$.}
\tablenotetext{b}{The blue color range of the TRGB with $1.5 \leq (V-I)_{0,TRGB}\leq 1.9$ was used.}
\tablenotetext{c}{Denotes standard deviation.}
\label{tab_lmc}
\end{deluxetable*}

%%%%%%%%%%%%
% Fig 11
%%%%%%%%%%%%

\begin{figure}
\centering
\includegraphics[scale=0.9]{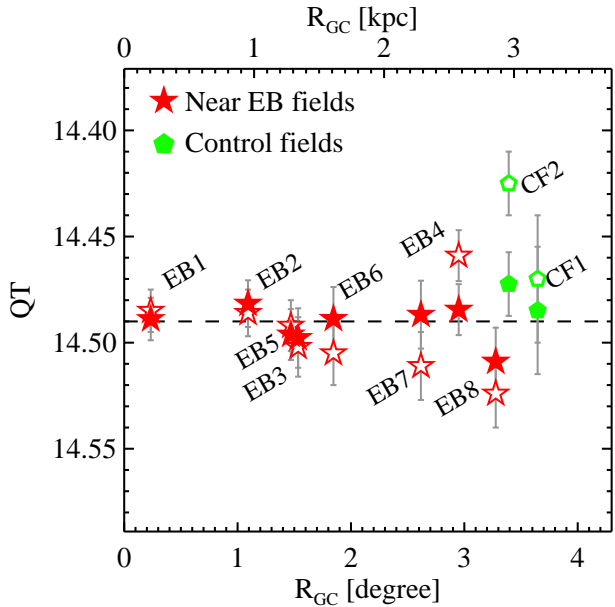} %white.eps}
\caption{
A comparison of the TRGB magnitudes from the fields around eclipsing binary stars (starlets) and from the control fields (pentagons) vs. galactocentric radius of the LMC.
The open and filled symbols denote the TRGB magnitudes before and after geometric distortion correction, respectively. 
The mean TRGB magnitude of eight eclipsing binary fields, corrected for the geometric distortion (filled stars), is estimated to be $QT=14.90\pm0.004$ (dashed line). %with a standard deviation of 0.024 mag (dashed line with shades).
}
\label{fig_compare_lmc}
\end{figure}

In {\color{red}{\bf Figure \ref{fig_cmd_lmc}}}, we plotted the extinction corrected $QT - (V-I)_0$ CMDs of eight fields around eclipsing binary stars (a-h) and two control fields (i, j). 
The extinction correction was done using the extinction map of the LMC given by \citet{has11}, who estimated  the extinction values from the mean color of red clump stars in the OGLE-III photometry. 
It is noted that we used the same extinction map as applied in \citet{pie13}.
All ten CMDs show various populations of stars including young main sequence stars (a vertical feature at $(V-I)_0\sim0.2$), blue and red helium burning stars (vertical and slanted features at $(V-I)_0\sim0.1$ and 1.2, respectively), old RGB stars (a dominant broad  slanted feature at $(V-I)_0\sim1.8$) and AGB stars (a broad slanted feature above the RGB) in the LMC.
The Milky Way foreground stars are also seen (vertical feature at $(V-I)_0\sim0.7$).
We measured the TRGB using the stars located in the shaded regions in {\color{red}{\bf Figure \ref{fig_cmd_lmc}}}, which were designed to avoid young stellar populations and to sample RGB stars as many as possible.
We plotted the edge detection responses as curved lines in each panel. All ten CMDs show strong peaks at $QT\sim14.5$ mag. 
We obtained quantitative values of the TRGB magnitudes and corresponding errors from the bootstrap resampling of ten thousand simulations. 
We also obtained the color dependence uncorrected TRGB magnitude, $I_{0,TRGB}$, by applying the same TRGB analysis to $I-(V-I)_0$ CMDs of the ten fields. Only the blue TRGB stars ($1.5\leq(V-I)_0\leq1.9$) showing a weak color dependence of the TRGB were used for the detection.
A summary of the $QT$ and $I$ magnitudes of the TRGB in each field of the LMC is listed in {\color{red}{\bf Table \ref{tab_lmc}}}.

%%%%%%%%%%%%
% Fig 12
%%%%%%%%%%%%

\begin{figure}
\centering
\includegraphics[scale=0.9]{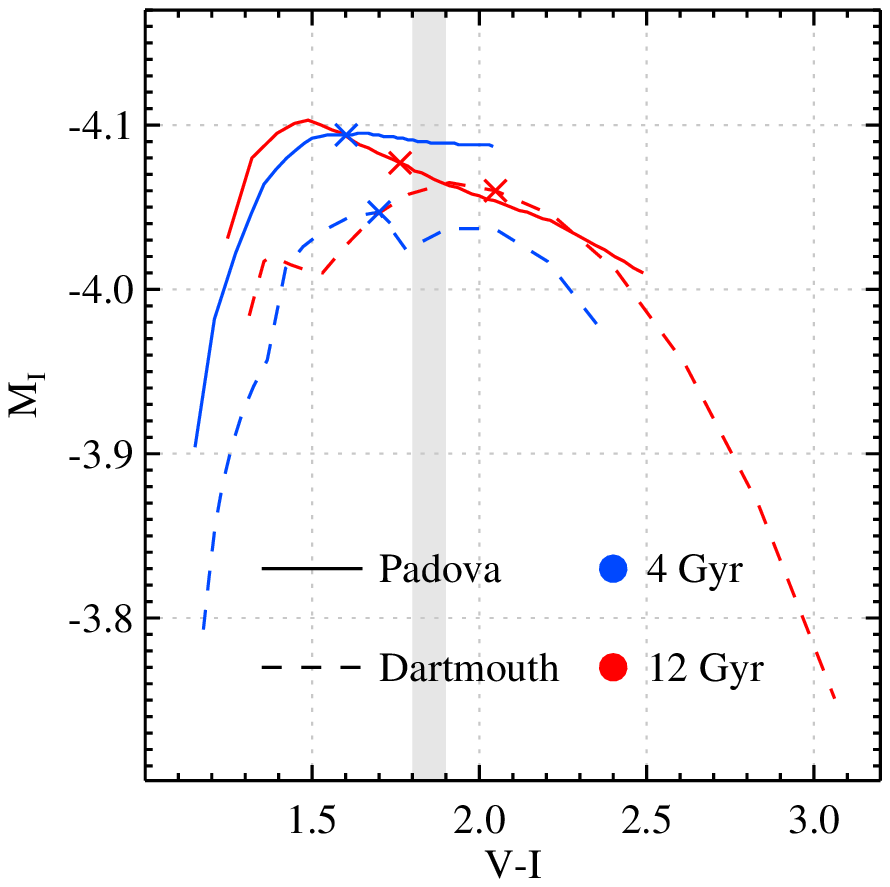} %white.eps}
\caption{
%\color{red}{
The TRGB for 4 Gyr (blue lines) and 12 Gyr (red lines) populations from the Padova \citep{gir00, mar13} (solid lines) and the Dartmouth \citep{dot08} (dashed lines) models.
Adopted metallicity range is $-2.3<[Fe/H]\leq-0.5$.
$[Fe/H]=-1.0$ points are marked by crosses.
A grey vertical region at $(V-I)=1.9$ indicates the mean TRGB color of the LMC.
Note that the TRGB for 4 Gyr Padova model at $(V-I)=1.9$ is $\sim0.03$ mag brighter than that of 12 Gyr model, whereas, those of Dartmouth models is $\sim0.03$ mag fainter.
%}
}
\label{fig_trgb_age}
\end{figure}

{\color{red}{\bf Figure \ref{fig_compare_lmc}}} shows a comparison of the TRGB magnitudes in the $QT$ system of eight eclipsing binary fields (starlets) and two control fields (pentagons) in the LMC. We corrected geometric distortion of the LMC by assuming the LMC disk  with an inclination angle of $28^\circ$ and a position angle of $128^\circ$, as assumed in \citet{pie13}. We plotted the TRGB magnitudes before and after correction of geometric distortion in {\color{red}{\bf Figure \ref{fig_compare_lmc}}} by open and filled symbols, respectively. 
The TRGB magnitudes are ranging from $QT \approx 14.425$ to 14.524 mag.
A weighted mean value of the TRGB magnitudes of all ten fields is $QT= 14.485\pm0.004$ mag with a standard deviation of 0.030 mag.
Similar values of $QT= 14.491\pm0.004$ mag and $QT= 14.432\pm0.013$ mag are derived from the eight eclipsing binary fields and two control fields, respectively.
Distortion correction contributes a very small change to the TRGB magnitudes: $0.004$ mag for all ten fields, $0.043$ mag for the two control fields, and $0.001$ mag for the eight eclipsing binary fields.
Similarly, distortion correction changes $I$-band TRGB magnitude of 0.007 mag for all ten fields, 0.034 mag for the two control fields, and 0.001 mag for the eight eclipsing binary fields ({\color{red}{\bf Table \ref{tab_lmc}}}).
Thus geometric distortion of the LMC is not significant in the TRGB estimation  in this study.

%{\color{red}{
The eight eclipsing binary fields used in this study cover the bar and disk of the LMC, so that we need to check the effect of composite stellar populations in the TRGB estimation.
\citet{sal05} conducted an analysis of artificial synthesis populations based on star-formation history of the LMC derived by \citet{hol99} and showed that
intermediate-age stars ($\sim$4 Gyr) can contaminate the RGB, resulting in wrong measurement of metallicity from colors by d[Fe/H]=0.5 dex.
They also pointed out that the metallicity difference of 0.5 dex may lead to 0.1 mag difference for the bolometric magnitude and corresponding $I$-band magnitude of the TRGB, when they use the TRGB calibration given in \citet{lee93}, $M_{bol,TRGB}$ = -0.19[Fe/H] - 3.81 mag.
However, they didn't show how much the TRGB magnitudes vary with different ages from their artificial CMD.

We compared the TRGB magnitudes for the old (12 Gyr) and the intermediate age (4 Gyr) populations using stellar isochrones provided by the Padova \citep{gir00,mar13} and Dartmouth \citep{dot08} groups and plotted them in {\color{red}{\bf Figure \ref{fig_trgb_age}}}. 
At the mean TRGB color of the LMC, $(V-I)= 1.8\sim1.9$, measured from our study, the intermediate age population from the Padova group (blue solid line) shows $0.02$ mag brighter magnitude than that of the old population (red solid line).
However, those of the Dartmouth group show an inverse trend: the intermediate age population is  $0.03$ mag fainter than that of the old population. 
This result implies that the the effect of the intermediate age population for the TRGB measurement is probably smaller than $0.03$ mag.

Two control fields used in this study, CF1 and CF2 shown in Figure \ref{fig_cmd_lmc} (i and j) are located far from the LMC bar, showing  no evidence of young and intermediate age population in the CMD. They are dominated by old halo stars. 
Measured $QT$ magnitudes of these two control fields are, on average, not different from those of 
other eight fields in the LMC disk and bar.
The mean differences between two control fields and the other eight fields are $0.015$ mag in $QT$ and 0.020 mag in $I$ after geometric correction.
From these results, we quote 0.02 mag as a systematic uncertainty for the mixture population of the LMC.

%}}

We adopted the distortion corrected mean TRGB magnitude of eight eclipsing binary fields, 
$QT=14.490$ mag and $I=14.524$, as mean values for TRGB magnitudes of the LMC. Conservative errors, 0.020 mag and 0.042 mag, which are standard deviations of these fields before distortion correction, is adopted as measurement errors.
Finally we obtain the zero-points of the TRGB: 
$M_{QT,TRGB}=-4.004\pm0.096$ mag for the $QT$ calibration and $M_{I,TRGB}=-3.970\pm0.102$ mag for the blue $I$ calibration.
%{\color{red}
Quoted uncertainties, 0.096 mag and 0.102 mag, are derived as summarized in {\color{red}{\bf Table \ref{tab_error}}}:
TRGB detection errors (0.020 mag and 0.042 mag),
an extinction correction error  (0.07 mag \citep{has11,gor16}),
a photometric zero-point error (0.03 mag), 
the $QT$ transformation error (0.014 mag), 
and the LMC distance  error (0.049 mag). 
The systematic uncertainty is dominated by the extinction error and the LMC distance error.
%}

% 18.477\pm 0.011\pm0.033 Freedman 12
% 18.494\pm 0.008\pm0.048 

% 0.071, 0.063 -> 0.047

\section{Summary of the Revised TRGB Calibration}

%%%%%%%%%%%%%%%%%%%%%%%%%%%%%%%%%%%%%%%%%%%
%% TABLES 6
%%%%%%%%%%%%%%%%%%%%%%%%%%%%%%%%%
\begin{deluxetable*}{llccccccc}
\tabletypesize{\scriptsize}
%\tabletypesize{\tiny}
\setlength{\tabcolsep}{0.05in}
%\rotate
\tablecaption{A Summary of the TRGB Calibrations in This Study}
\tablewidth{0pt}
%\tablehead{\colhead{System} & \colhead{$Color$} & \colhead{$\alpha$} &  \colhead{$\beta$} & \colhead{$\gamma$} & \colhead{Color range}& \multicolumn{3}{c}{Zero-point} \\
%&  \multicolumn{4}{r}{$QT = I(\rm F814W)_0 - \alpha(Color-\gamma)^2 - \beta(Color-\gamma)$} &  & NGC 4258 & LMC & N4258 + LMC 
\tablehead{ & \multicolumn{4}{r}{$QT = I(\rm F814W)_0 - \alpha(Color-\gamma)^2 - \beta(Color-\gamma)$} & & \multicolumn{3}{c}{Zero-point}\\
\colhead{System} & \colhead{$Color$} & \colhead{$\alpha$} &  \colhead{$\beta$} & \colhead{$\gamma$} & \colhead{Color range}& NGC 4258 & LMC & N4258 + LMC 
%\multicolumn{3}{l}{$QT$ calibration}
}
\startdata

\multicolumn{3}{l}{$QT$ calibration} \\
ACS/WFC 	& $(\rm F606W-F814W)_0$ & $0.159\pm0.010$ & $-0.047\pm0.020$ & $1.1$ & ... & $-4.021\pm0.067$ & $-4.002\pm0.101$ & $-4.015\pm0.056$\\
ACS/WFC 	& $(\rm F555W-F814W)_0$ & $0.116\pm0.007$ & $-0.043\pm0.017$ & $1.6$ & ... & $-4.017\pm0.067$ & $-3.998\pm0.101$ & $-4.011\pm0.056$\\
J.C.		& $(V-I)_0$ 		& $0.091\pm0.006$ & $-0.007\pm0.013$ & $1.5$ & ... & $-4.023\pm0.073$ & $-4.004\pm0.096$ & $-4.016\pm0.058$\\
WFC3/UVIS	& $(\rm F606W-F814W)_0$ & $0.150\pm0.010$ & $-0.050\pm0.023$ & $1.1$ & ... & $-4.031\pm0.068$ & $-4.012\pm0.101$ & $-4.025\pm0.057$\\
WFC3/UVIS 	& $(\rm F555W-F814W)_0$ & $0.113\pm0.007$ & $-0.048\pm0.019$ & $1.6$ & ... & $-4.026\pm0.068$ & $-4.007\pm0.101$ & $-4.020\pm0.057$\\
WFPC2		& $(\rm F606W-F814W)_0$ & $0.179\pm0.012$ & $-0.034\pm0.024$ & $1.0$ & ... & $-3.992\pm0.125$ & $-3.973\pm0.139$ & $-3.984\pm0.121$\\
WFPC2 		& $(\rm F555W-F814W)_0$ & $0.123\pm0.008$ & $-0.051\pm0.020$ & $1.5$ & ... & $-4.014\pm0.069$ & $-3.995\pm0.092$ & $-4.007\pm0.060$\\

\hline
\multicolumn{3}{l}{Blue $I$ calibration}\\
ACS/WFC 	& $(\rm F606W-F814W)_0$ & ... & ... & ... &Color $\leq1.5$ & $-4.034\pm0.068$ & $-3.968\pm0.106$ & $-4.015\pm0.057$\\
ACS/WFC 	& $(\rm F555W-F814W)_0$ & ... & ... & ... &Color $\leq2.1$ & $-4.030\pm0.068$ & $-3.964\pm0.106$ & $-4.008\pm0.057$\\
J.C.		& $(V-I)_0$ 			& ... & ... & ... &Color $\leq1.9$ & $-4.036\pm0.074$ & $-3.970\pm0.102$ & $-4.013\pm0.060$\\
WFC3/UVIS	& $(\rm F606W-F814W)_0$ & ... & ... & ... &Color $\leq1.5$ & $-4.044\pm0.069$ & $-3.978\pm0.107$ & $-4.025\pm0.058$\\
WFC3/UVIS 	& $(\rm F555W-F814W)_0$ & ... & ... & ... &Color $\leq2.1$ & $-4.039\pm0.069$ & $-3.973\pm0.107$ & $-4.020\pm0.058$\\
WFPC2		& $(\rm F606W-F814W)_0$ & ... & ... & ... &Color $\leq1.5$ & $-4.005\pm0.126$ & $-3.939\pm0.143$ & $-3.976\pm0.121$\\
WFPC2 		& $(\rm F555W-F814W)_0$ & ... & ... & ... &Color $\leq1.9$ & $-4.027\pm0.069$ & $-3.961\pm0.098$ & $-4.005\pm0.062$\\

\enddata
\label{tab_cal}
\end{deluxetable*}

%%%%%%%%%%%%%%%%%%%%%%%%%%%%%%%%%%%%%%%%%%%
%% TABLE 7
%%%%%%%%%%%%%%%%%%%%%%%%%%%%%%%%%
\begin{deluxetable*}{llllcccc}
\tabletypesize{\scriptsize}
%\tabletypesize{\tiny}
\setlength{\tabcolsep}{0.05in}
%\rotate
\tablecaption{A List of the Previous and Revised TRGB Calibrations}
\tablewidth{0pt}
\tablehead{ \colhead{Reference} & \colhead{Calibration} & \colhead{Zero-point} &   \colhead{Note} \\
& & \colhead{$M_{I,TRGB}$} & 
}
\startdata
\citet{lee93} & $-4.0\pm0.1$ & $-4.0\pm0.1$ & MW GCs\\
\citet{sal97} & $\rm -3.732+0.588[M/H]+0.193[M/H]^2$ & $-4.16$ & $\rm -2.35\leq [M/H]\leq-0.57$ \\ %, $[M/H]=-1.2$, Model\\
\citet{fer00} & $-4.06\pm0.07\pm0.13$ 		& $-4.06\pm0.07\pm0.13$ & Cepheids, $\gamma=0.00$ \\
\citet{fer00} & $-3.99\pm0.07\pm0.13$ 		& $-3.99\pm0.07\pm0.13$ & Cepheids, $\gamma=-0.24$\\
\citet{bel01} & $\rm -3.66+0.48[Fe/H]+0.14[Fe/H]^2$	& $-4.07\pm0.12$ & $\rm [Fe/H]=-1.6$, E.B, $\omega$ Cen \\%, New $M_{bol}$ + DA90 \\
\citet{bel04} & $\rm -3.629+0.679[M/H]+0.258[M/H]^2$	& $-4.07\pm0.12$ & $\rm [M/H]=-1.2$, E.B, $\omega$ Cen\\
\citet{riz07} & $-4.05+0.217[(V-I)_0-1.6]$			& $-4.07$ & Six galaxies \\
\citet{bel08}& $-3.939-0.194(V-I)_0+0.080(V-I)_0^2$& $-4.05$ &  \\
\citet{tam08} & $-4.05\pm0.02\pm0.10$ 		& $-4.05\pm0.02\pm0.10$ & RR Lyrae, LMC scale\\
\citet{mad09} & $-4.05+0.2[(V-I)_0-1.5]$	& $-4.05$ &  $T$ magnitude\\
\citet{fus12} & $-3.63-0.40Col^a+0.08Col^2$		& $-4.13\pm0.01$ & $\rm F814W$, model\\
 This study    & $-4.016+0.091((V-I)_0-1.5)^2 - 0.007((V-I)_0-1.5)$ & $-4.015\pm0.059$ & NGC 4258 and the LMC \\
%1  & \citet{sal97} & $-4.17$ & model\\
\hline
%1 & \multicolumn{2}{c}{Mean, standard deviation} & $-4.064, 0.049$
%1 & Straight mean & & $-4.064$ $(\sigma=0.049)$
\enddata
\tablenotetext{a}{Col $= (\rm F475W - F814W)_0$.}
\label{tab_cal_list}
\end{deluxetable*}

In {\color{red}{\bf Table \ref{tab_cal}}}, we summarize the two revised TRGB calibrations, the $QT$ and the blue $I$ calibrations, in several photometric systems. 
%{\color{red}{
Color dependence correction terms ($\alpha$, $\beta$, and $\gamma$) in these photometric systems were determined by applying the photometric transformation as described in Section 2.4. 
The same method was applied to obtain the zero-point values in each photometric system.
The zero-point uncertainties associated with the photometric transformation from F814W in ACS to other systems are estimated to be 
0.03 mag for the $I$, 
0.01 mag for both F555W--F814W and F606W--F814W combinations of F814W in WFC3, 
0.015 mag for F555W--F814W conbination of F814W in WFPC2, and 
0.106 mag for F606W--F814W combination of F814W in WFPC2 \citep{sir05, jan15}.
%}}

Zero-points from NGC 4258 and the LMC shows a good agreement.
A weighted mean of the zero-points from NGC 4258 and the LMC yields $M_{QT,TRGB}=-4.016\pm0.058$ and $M_{I,TRGB}=-4.013\pm0.060$ in  $I$-band. 
These values are similar to the values given in previous calibrations, $M_{I,TRGB} = -4.05\pm0.12$ \citep{bel01, bel04, riz07, mad09}, but its quoted error is two times smaller.

\section{Discussion}

\subsection{Comparison with Previous TRGB Calibrations}

%%%%%%%%%%%%
% Fig 13
%%%%%%%%%%%%

\begin{figure}
\centering
\includegraphics[scale=0.9]{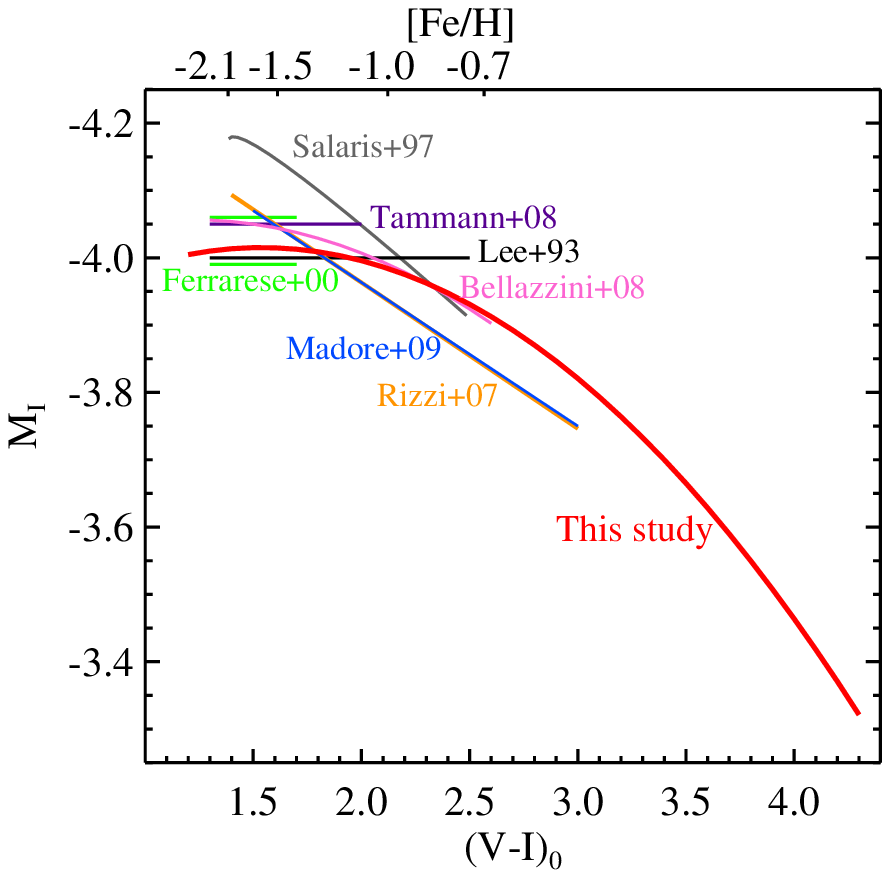} %white.eps}
\caption{A comparison of the TRGB calibrations in $M_I - (V-I)_0$ domain in this study and the previous studies \citep{lee93,sal97,fer00,riz07,bel08,tam08,mad09}.
}
\label{fig_com1}
\end{figure}

There have been several studies of the TRGB calibration including both color-magnitude dependence and zero-point calibrations as summarized in {\color{red}{\bf Table \ref{tab_cal_list}}}.
\citet{dac90} provided the bolometric magnitude and bolometric correction of the TRGB as a function of metallicity and color, based on $VI$ photometry of six globular clusters in the Milky Way. With these relations, \citet{lee93} estimated the absolute $I$-band magnitude of the TRGB, $M_{I,TRGB}=-4.0\pm0.1$ mag, for metal poor stellar systems ([Fe/H] $<-0.7$).
\citet{sal97} derived a TRGB calibration having quadratic terms of metallicity from their theoretical stellar models. Absolute $I$-band magnitude of the TRGB at the fiducial metallicity, [M/H] $=-1.2$, is estimated to be $-4.16$ mag, slightly brighter than that of \citet{lee93}. 
This discrepancy is coming from the different estimation of the TRGB bolometric luminosity \citep{fer00}.
\citet{fer00} calibrated the TRGB luminosity from the Cepheid distance estimations of nine galaxies in the Local Group. 
Absolute $I$-band magnitudes of the TRGB, $-3.99$ and $-4.06$ mag with an uncertainty of $\sim0.15$ mag, were derived depending on the adoption of metallicity dependence of Cepheid variables ($\gamma=\delta(m-M)_0/\delta [{\rm O/H}]=0.00$ and --0.24, respectively).

Later \citet{bel01, bel04} and \citet{bel08} presented another set of the TRGB calibration based on the larger sample of the Milky Way globular clusters:
$M_{\rm bol,TRGB} = -0.12 [{\rm Fe/H}] -3.76$ and
$M_{\rm I,TRGB} = 0.14 [{\rm Fe/H}]^2 + 0.48 [{\rm Fe/H}] -3.66$.
\citet{bel01} also presented another calibration based on $\omega$ Centauri for which a direct distance estimate based on one eclipsing binary star \citep{tho01} is available. Adopting a metallicity of [Fe/H]$=-1.7$ for $\omega$ Centauri, they derived $M_{\rm I,TRGB} = -4.04\pm0.12$ where the error is dominated by the distance measurement error for $\omega$ Centauri. This error has been used a value for the systematic error for the TRGB calibration \citep{mag08, con11, sha11, lee12, lee13, jan14, jan15, lee16}.
This value of the calibration is consistent with that from the quadratic calibration for [Fe/H] $=-1.7$ in this study, $M_{\rm I,TRGB} = -4.02$.

On the other hand \citet{riz07} provided a linear color dependence of the TRGB from the analysis of HST photometry data for six nearby galaxies:
$M_{I,TRGB} = -4.05 + 0.217 [(V-I)_0 -1.6]$. They constrained the zero-point by using five Local Group galaxies, whose distances are obtained from the horizontal branch stars. Adopting the calibration of the horizontal branch stars given by \citet{car00}, they derived a zero-point value, $-4.05\pm0.02$ mag at a fiducial color, $(V-I)_0=1.6$ (corresponding to $\rm [Fe/H]=-1.5)$. The systematic error of this value is mainly due to the uncertainty in the mean absolute magnitude of the horizontal branch stars, which is
$\sim 0.1$ mag \citep{car00}.

\citet{tam08} compiled the data for 24 galaxies that have estimates for both the TRGB magnitudes and the RR Lyrae distances in the literature, and derived an absolute magnitude of the TRGB, $M_{I,TRGB}=-4.05$ mag with a mean error of 0.02 mag.
The systematic error of this value is mainly due to the uncertainty in the mean magnitude of the RR Lyrae, which is
$\sim 0.1$ mag \citep{pop98a, pop98b}.

\citet{mad09} described a novel approach for a robust determination of the TRGB magnitude by introducing a new magnitude system, $T$ magnitude, which is a color (metallicity) corrected $I$-band magnitude:
$T=I_0-\beta((V-I)_0-1.5)$. They
adopted a value for the slope, $\beta=0.20\pm0.05$ for $1.5<(V-I)_0<3.0$, 
 from a linear approximation of the color-luminosity relation given by \citet{bel01, bel04}. 
They showed  that the $T$ magnitude works well in the case 
of NGC 4258 \citep{mag08, mad09}.

Recently \citet{fus12} carried out a TRGB calibration with the BaSTI stellar models  \citep{pie04, pie06}, updating the conductive opacity evaluations provided by \citet{cas07}.
They presented variations of the TRGB magnitude as a function of metallicity and ${\rm F475W - F814W}$ color in their Figure 4. At the fiducial metallicity ([Fe/H]$=-1.6$), their calibration yields a value, $M_{\rm{F814W},TRGB}=-4.13$, which is slightly brighter than other calibrations.

%%%%%%%%%%%%
% Fig 14
%%%%%%%%%%%%

\begin{figure}
\centering
\includegraphics[scale=0.8]{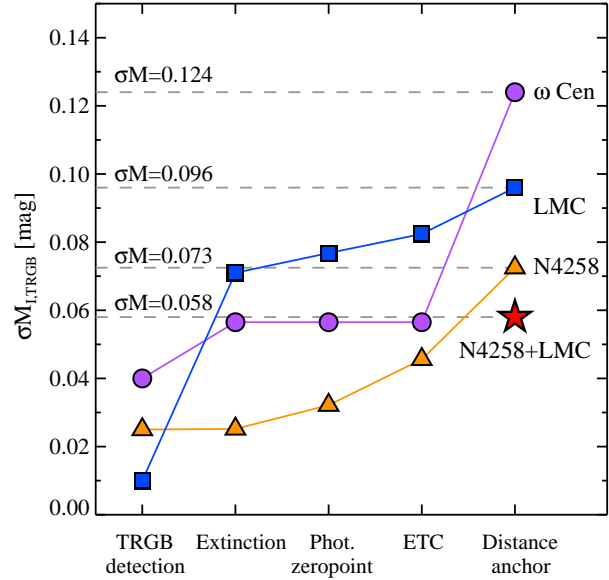} %white.eps}
\caption{
Propagation of uncertainties in the zero-point calibration of the TRGB.
Note that the previous calibration based on $\omega$ Centauri gives a zero-point uncertainty of 0.124 mag \citep{bel01, bel04}.
When NGC 4258 and LMC are used as a distance anchor, uncertainties are estimated to be 0.073 and 0.096 mag, respectively. 
Combining these two anchors (NGC 4258 and LMC) provides the most accurate calibration with an uncertainty of 0.058 mag (red starlet), 53\% smaller than the value given by \citet{bel01, bel04}.
Individual error values are listed in {\color{red} \bf Table \ref{tab_error}}.
}
\label{fig_sys}
\end{figure}

In {\color{red}{\bf Figure \ref{fig_com1}}}, we compare our TRGB calibration with the previous ones %the TRGB calibrations 
in the $M_I - (V-I)_0$ domain. 
In the figure we set line lengths to be consistent with the color range used for each TRGB calibration.
In the upper axis of the figure we also marked the metallicity values of the TRGB using the relation between $(V-I)_0$ and [Fe/H] for [Fe/H]$<-0.7$, %derived from the data for the Milky Way globular clusters in the sample of \citet{dac90} 
given by \citet{bel01} ($(V-I)_{0,TRGB} = 0.581  [{\rm Fe/H}]^2 + 2.742 [{\rm Fe/H}] +4.013 $).
Several important features are seen in this figure.
First, the revised TRGB calibration derived in this study (red line) covers a much wider color range, $1.2 \lesssim(V-I)_0 \lesssim4.3$, than other calibrations.  
Second, our TRGB calibration shows a color-magnitude relation that is relatively flat at the blue end and steep at the red color range. %, respectively. 
Some of the previous calibrations adopted either a flat relation \citep{lee93, fer00, tam08} or color-dependent relations  \citep{mad09, riz07, sal97}. 
Third, the color dependence at the color range of $2.0\lesssim(V-I)_0\lesssim3.0$ derived in this study is similar to the results given by  \citet{sal97}, \citet{mad09}, \citet{riz07}, and \citet{bel08}.
Fourth, our calibration shows an absolute magnitude of the TRGB, $M_I=-4.016\pm0.058$, at a fiducial color, $(V-I)_0=1.5$. 
It agrees well with the values from all of the previous calibrations within uncertainties.

{\color{red}{\bf Figure \ref{fig_sys}}} shows the propagation of the zero-point uncertainties in the previous calibration based on $\omega$ Centauri (circles, \citet{bel01,bel04}) and the revised $QT$ calibrations based on NGC 4258 and the LMC in this study (triangles and squares). 
Four main sources of uncertainties (TRGB detection, extinction, photometric zero-point, and distance anchor uncertainties) and additional sources of uncertainties (ETC, which includes color transformation, photometric transformation, $QT$ transformation uncertainties) listed in {\color{red} \bf Table \ref{tab_error}} are indicated.
We calculated cumulative errors of each source (a quadratic sum of uncertainties including all the previous sources), according to this order. 
\citet{bel01, bel04} presented a luminosity calibration of the TRGB 
based on the distance to $\omega$ Centauri,
and derived a total uncertainty, 0.124 mag from the quadratic sum of three uncertainties: 0.04, 0.04, and 0.11 mag for extinction, TRGB detection, and distance anchor uncertainties, respectively.
They did not provide any uncertainty for the photometric zero-point.
On the other hand, the revised TRGB calibration in this study yields total systematic uncertainty of 0.073 mag from NGC 4258 and 0.096 mag from the LMC as described in Section 3. 
A weighted mean of these two calibrations gives systematic uncertainty of 0.058 mag, which is about a half of the uncertainty given by \citet{bel01, bel04}.
This value represents the systematic error of the revised TRGB calibration in this study.

\subsection{Comparison with Stellar Isochrones}

%%%%%%%%%%%%
% Fig 15
%%%%%%%%%%%%

\begin{figure}
\centering
\includegraphics[scale=0.9]{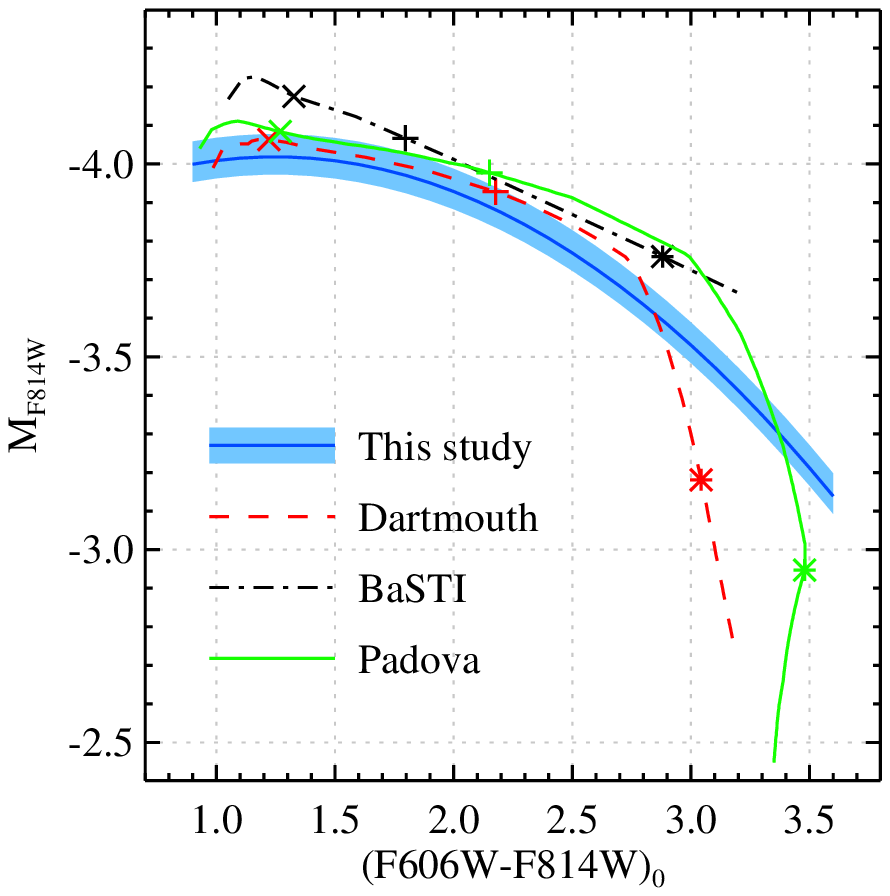} %white.eps}
\caption{A comparison of the revised TRGB calibration (blue solid line with shaded region) with the TRGB from 12 Gyr stellar models: Dartmouth \citep{dot08} (red dashed line), BasTI \citep{pie04, pie06} (black dot-dashed line), and Padova \citep{gir00, mar13} (green solid line). Cross, plus, and asterisk symbols in each stellar model indicate [M/H]$=-1.0, -0.5,$ and $0.0$, respectively.  The shaded region indicates a zero-point uncertainty, 0.046 mag, in this study.
}
\label{fig_com2}
\end{figure}

In {\color{red}{\bf Figure \ref{fig_com2}}} we compare the empirical calibration of the TRGB (solid line with shaded region) derived in this study with the TRGB from three theoretical stellar models: Dartmouth \citep{dot08} (dashed line), BasTI \citep{pie04, pie06} (dot-dashed line) , and Padova \citep{gir00, mar13} (solid line).  We set age of 12 Gyr and metallicity range of [M/H] = --2.2 to +0.2 for stellar models. A fixed alpha element abundance, $[\alpha/Fe]=+0.2$, was adopted for the Dartmouth model. 

A few notable features are seen in this figure.
First, the blue color range ($(\rm F606W-F814W)_0\lesssim1.6$) of the empirical calibration shows a good agreement with those of the Dartmouth and Padova models.
In the case of the BasTI model, however, it shows $0.20 \sim 0.15$ mag systematic offset and is steeper than those of other calibrations.
Second, all four calibrations show a good agreement at $(\rm F606W-F814W)_0\sim2.0$, which  corresponds to [M/H]$\sim-0.5$.
Third, a significant dispersion is seen at the red color range ($(\rm F606W-F814W)_0\gtrsim2.7$).
The BasTI model gives relatively bright TRGB luminosity and a shallower color dependence. Dartmouth and Padova models show much steeper than others.
Our empirical calibration of the TRGB will be very useful for improving the modelling of stellar evolution in the future.

\subsection{Implications on the Distance Scale and the Hubble Constant}

We obtained the zero-point of the TRGB, $M_{QT,TRGB}=-4.023\pm0.073$ from NGC 4258 and $M_{QT,TRGB}=-4.004\pm0.096$ from the LMC. The zero-point difference is 
 $\Delta M_{QT,TRGB}=-0.019\pm0.121$. Thus, the two zero-point values agree very well. 
From the TRGB magnitudes of NGC 4258 and the LMC, the distance modulus difference between two galaxies is obtained, $\Delta(m-M)_0=10.874\pm0.083$.
This value shows excellent agreement with the value from the Cepheid variables, $\Delta(m-M)_0=10.88\pm0.04_r\pm0.05_s$ \citep{mac06}.
If we adopt the TRGB zero-point from the LMC, the TRGB distance to NGC 4258 is estimated to be $(m-M)_0=29.368\pm0.023_r\pm0.096_s$, consistent with the recent megamaser distance to NGC 4258, $(m-M)_0=29.404\pm0.049_r\pm0.043_s$ \citep{hum13} and $(m-M)_0=29.387\pm0.049_r\pm0.029_s$ \citep{rie16}.

The revised TRGB calibration derived in this study enables to determine more precise TRGB distances to nearby galaxies and a corresponding value of the Hubble constant.
There have been several studies to determine the Hubble constant based on the TRGB \citep{fer00,  mou09a, mou09b, tam08, his11, lee12, tam13, lee13, jan15}. 
Most of these studies were based on the zero-point of the TRGB, $M_I=-4.0\sim-4.1$ mag with a systematic uncertainty of $0.10 \sim 0.12$ mag. The revised TRGB calibration combined with both NGC 4258 and LMC anchors yields a TRGB zero-point of  $M_QT=-4.016\pm0.058$ mag, slightly fainter and much more accurate than those of previous calibrations.
Thus, the value of the Hubble constant derived in the previous studies can be slightly increased if the revised TRGB calibration is used.

Our previous studies, \citet{lee12, lee13} and \citet{jan15}, presented a value of the Hubble constant, 
$H_0=69.8\pm2.6\rm(random)\pm3.9 \rm(systematic)$ \kmsMpc~ and 
$H_0=72.2\pm3.3\rm(random)\pm4.0 \rm(systematic)$ \kmsMpc~ based on five and three  TRGB calibrated SNe Ia. 
In these studies we used the old TRGB calibration given by \citet{riz07} and the systematic error of 0.12 mag given by \citet{bel01}. 
If we apply the revised TRGB calibration including the $QT$ magnitude and updated zero-point ($M_{QT,TRGB}=-4.016\pm0.058$ mag), then the value of $H_0$ would be increased by $\sim1$ \kmsMpc~ and the systematic uncertainty would be decreased by a factor of 2. 
A detailed analysis of the TRGB distances to eight SN Ia host galaxies and the Hubble constant based on this new calibration will be presented in  our upcoming paper (Jang\&Lee 2016, manuscript submitted).

\section{Summary and Conclusion}

The TRGB has been used as a reliable distance indicator for resolved stellar systems.
However, its calibration needs to be improved to derive more precise distances to nearby galaxies and corresponding value of the Hubble constant. 
We present a revised TRGB calibration including the color dependence and zero-point calibrations, accurate to 0.058 mag (2.7\% of distance). Primary results are as follows.

\begin{itemize}
\item We obtained deep photometry of the resolved stars in eight nearby galaxies from the archival $HST/ACS$ images. By applying quantitative TRGB detections on $\rm F814W_0-(F606W-F814W)_0$ CMDs, we derived color-luminosity relation of the TRGB, which is described by a quadratic equation.

\item From the photometry of two nearby globular clusters (NGC 2419 and 47 Tuc), we derived color transformation between $\rm F555W-F814W$ and $\rm F606W-F814W$ in ACS/WFC system.

\item 
The zero-point of the TRGB was determined from the photometry of two nearby galaxies, NGC 4258 and the LMC, to which geometric distances are known. \\

\item 
We carried out PSF photometry using five different reduction methods on the HST/ACS field of NGC 4258 and compared the results. We found that output magnitudes agree well within 0.06 mag at the TRGB magnitude level. Especially, F814W band photometry shows excellent agreement within 0.02 mag, regardless of reduction methods.

\item
We provide the revised TRGB calibration in several filter systems including Johnson-Cousins, ACS/WFC, WFC3/UVIS, and WFPC2, as listed in {\color{red}{\bf Table \ref{tab_cal}}}.

\end{itemize}

\bigskip

The authors thank the anonymous referee for the useful comments that improved the original manuscript.
I.S.Jang thank Maria-Rosa L. Cioni and Jesper Storm for helpful discussions and comments.
This work was supported by the National Research Foundation of Korea (NRF) grant funded by the Korea Government (MSIP) (No. 2012R1A4A1028713). %BRL (No. 2013...). % Core
This paper is based on image data obtained from the Multimission Archive at the Space Telescope Science Institute (MAST).

\end{document}